\documentclass[aps,showpacs,prb,floatfix]{revtex4}
\usepackage{amsmath,amssymb,graphicx,bm,epsfig}

\input{tcilatex}

\begin{document}

\title{Electronic Structure and Magnetic Properties of Solids}
\author{S. Y. Savrasov$^{a}$, A. Toropova$^{b},$M. I. Katsnelson$^{c},$ A.
I. Lichtenstein$^{d}$, V. Antropov$^e$, G. Kotliar$^{b}$ }
\affiliation{$^{a}$Department of Physics, New Jersey Institute of Technology, Newark, NJ
07102, USA}
\affiliation{$^{b}$Department of Physics and Center for Material Theory, Rutgers
University, Pscataway, NJ 08854, USA}
\affiliation{$^{c}$Department of Physics, University of Nijmegen, Nijmegen, The
Netherlands}
\affiliation{$^{d}$Department of Physics, Hamburg University, Hamburg, Germany}
\affiliation{$^{e}$Condensed Matter Physics Department, Ames Laboratory, Ames, IA 50011,
USA}

\begin{abstract}
We review basic computational techniques for simulations of various magnetic
properties of solids. Several applications to compute magnetic anisotropy
energy, spin wave spectra, magnetic susceptibilities and temperature
dependent magnetisations for a number of real systems are presented for
illustrative purposes.
\end{abstract}

\date{May 2003}
\maketitle

\section{Introduction}

This review covers main techniques and their applications developed in the
past to calculate properties of magnetic systems by the methods of
electronic structure theory of solids. The basic tool which was used in
connections with these developments is a spin dependent version of density
functional theory \cite{LDA} which will be reviewed in Section II in its
most general non--collinear form and including so called LDA+U method \cite%
{LDA+U}. One of the oldest application of this method is the problem of
magnetic anisotropy energy (MAE). Due to its technological relevance as well
as the smallness of MAE, such calculations represent a real challenge to the
theory and we review these efforts in Section III. In Section IV we address
the problem of computing non--collinear spin alignments which is solved
using an elegant spin spiral approach \cite{SpinSpiral}. This allows us to
simulate adiabatic spin dynamics of real magnets with relatively cheap
computational effort, and some of the calculations will be discussed and
compared to experiments. Another development which has been undertaken in
the past is the direct calculation of exchange integrals by utilizing a
formula of linear response theory \cite{Licht}. This is reviewed in Section
V. The spin waves, their dispersions and lifetimes as well as other spin
fluctuations formally appear in the structure of dynamical spin
susceptibility. The developments of electronic structure based on linear
response approaches which access this quantity is reviewed in Section VI.
The effect of electronic correlations are not small in the real materials
and static mean field treatment done with local spin density functional or
LSDA+U method may not be adequate for a whole range of solids. Bringing the
effects of dynamical correlations to compute magnetic properties of real
systems has been recently undertaken by utilizing the dynamical mean field
theory \cite{DMFTReview}. This approach will be reviewed in Section VII
together with some of its most recent applications. The reader is also
refereed to several other reviews \cite{JMMAntropov,PhysRepSan} on a similar
subject as well as a few excellent books \cite{Moria,Kubler}.

\section{Spin Density Functionals in Non-Collinear Form}

Here we describe a basic tool for computational studies of magnetic
phenomena, a spin dependent version of the density functional theory \cite%
{LDA}. We use general non--collinear notations and also treat the effects of
spin --orbit coupling. We utilize local spin density approximation (LSDA)\
to the exchange--correlation functional and also show how the corrections
due to strong electronic correlation effects appear in the functional in its
simplest Hartree--Fock form known as the LDA+U method \cite{LDA+U}. In
section V we will also describe a more general first--principle method based
on dynamical mean field theory \cite{DMFTReview} to include the correlation
effects in a more rigorous manner.

To describe a magnetic solid we consider a system of fermions under an
external potential $V_{ext}$ and an external magnetic field $\mathbf{B}%
_{ext} $. It is useful to introduce the notion of the Kohn-Sham potential $%
V_{eff}(\mathbf{r})$ and the Kohn--Sham magnetic field $\mathbf{B}_{eff}(%
\mathbf{r})$. When spin--orbit coupling is present, the intra--atomic
magnetization $\mathbf{m}(\mathbf{r})$ is not collinear, and the solid may
choose different atom dependent quantization axes which makes magnetic
moments pointing in different directions. Therefore, the magnetization must
be treated as a general vector field, which also realizes non--collinear
intra--atomic nature of this quantity. Such general magnetization scheme has
been recently discussed \cite{singh}. In the non--collinear version of LSDA
approach the total energy functional $E$ is considered as a functional of
two variables the charge density $\rho (\mathbf{r})$ and magnetization
density $\mathbf{m}(\mathbf{r})$%
\begin{equation}
\rho (\mathbf{r})=\sum_{\mathbf{k}j}f_{\mathbf{k}j}\{\vec{\psi}_{\mathbf{k}%
j}(\mathbf{r})|\hat{I}|\vec{\psi}_{\mathbf{k}j}(\mathbf{r})\}=\sum_{\mathbf{k%
}j}f_{\mathbf{k}j}\sum_{\sigma =\uparrow \downarrow }\psi _{\mathbf{k}%
j}^{(\sigma )\ast }(\mathbf{r})\psi _{\mathbf{k}j}^{(\sigma )}(\mathbf{r})
\label{SDFrho}
\end{equation}%
\begin{equation}
\mathbf{m}(\mathbf{r})=g\mu _{B}\sum_{\mathbf{k}j}f_{\mathbf{k}j}\{\vec{\psi}%
_{\mathbf{k}j}(\mathbf{r})|\mathbf{\hat{s}}|\vec{\psi}_{\mathbf{k}j}(\mathbf{%
r})\}=g\mu _{B}\sum_{\mathbf{k}j}f_{\mathbf{k}j}\sum_{\sigma \sigma ^{\prime
}=\uparrow \downarrow }\psi _{\mathbf{k}j}^{(\sigma )\ast }(\mathbf{r})%
\mathbf{\hat{s}}_{\sigma \sigma ^{\prime }}\psi _{\mathbf{k}j}^{(\sigma
^{\prime })}(\mathbf{r})  \label{SDFmag}
\end{equation}%
where $\left\{ \left\vert {}\right\vert \right\} $ denotes averaging over
spin degrees of freedom only, the spin angular momentum operator is
expressed in terms of Pauli matrices $\mathbf{\hat{s}}=\mathbf{\hat{\sigma}}%
/2,$ $\hat{I}$ is $2\times 2$ unit matrix, $g$ is the gyromagnetic ratio
which is for electrons equal to 2, and $\vec{\psi}_{\mathbf{k}j}(\mathbf{r})$
are the non--interacting Kohn--Sham particles are formally described by the
two--component spinor wave functions 
\begin{equation}
\vec{\psi}_{\mathbf{k}j}(\mathbf{r})=\left( 
\begin{array}{c}
\psi _{\mathbf{k}j}^{(\uparrow )}(\mathbf{r}) \\ 
\psi _{\mathbf{k}j}^{(\downarrow )}(\mathbf{r})%
\end{array}%
\right)  \label{SDFpsi}
\end{equation}%
which define the charge and non--collinear magnetization densities of the
electrons.

LSDA$+$U method \cite{LDA+U} introduces additional variable "occupancy spin
density matrix" $\hat{n}_{ab}$%
\begin{equation}
\hat{n}_{ab}=\left( 
\begin{array}{cc}
n_{ab}^{\uparrow \uparrow } & n_{ab}^{\uparrow \downarrow } \\ 
n_{ab}^{\downarrow \uparrow } & n_{ab}^{\downarrow \downarrow }%
\end{array}%
\right)  \label{SDFnab}
\end{equation}%
which represents the correlated part of electron density. To build up
occupancy matrix one introduces a set of localized orbitals $\phi _{a}(%
\mathbf{r})$, associated with correlated electrons. Then 
\begin{equation}
n_{ab}^{\sigma \sigma ^{\prime }}=\sum_{\mathbf{k}j}f(\epsilon _{\mathbf{k}%
j})\langle \psi _{\mathbf{k}j}^{(\sigma )}|\phi _{a}\rangle \langle \phi
_{b}|\psi _{\mathbf{k}j}^{(\sigma ^{\prime })}\rangle  \label{SDFmat}
\end{equation}%
The occupancy matrix becomes non-diagonal in spin space when spin--orbit
coupling is taken into account. We include spin--orbit coupling in a
variational way as suggested by Andersen~\cite{OKA}.

One introduces a Lagrange multipliers matrix $\Delta \hat{V}_{ab}$ to
enforce~(\ref{SDFmat}). The LSDA$+$U total energy functional is given by: 
\begin{eqnarray}
&&E[\rho (\mathbf{r}),\mathbf{m}(\mathbf{r}),\hat{n}_{ab}]=\sum_{\mathbf{k}%
j}f(\epsilon _{\mathbf{k}j})\epsilon _{\mathbf{k}j}-\int V_{eff}(\mathbf{r}%
)\rho (\mathbf{r})d\mathbf{r}+\int \mathbf{B}_{eff}(\mathbf{r})\mathbf{m}(%
\mathbf{r})d\mathbf{r-}\sum_{ab}\sum_{\sigma \sigma ^{\prime }}\Delta
V_{ab}^{\sigma \sigma ^{\prime }}n_{ab}^{\sigma \sigma ^{\prime }}  \notag \\
&&+\int V_{ext}(\mathbf{r})\rho (\mathbf{r})d\mathbf{r}-\int \mathbf{B}%
_{ext}(\mathbf{r})\mathbf{m}(\mathbf{r})d\mathbf{r}+\frac{1}{2}\int d\mathbf{%
r}d\mathbf{r}^{\prime }\frac{\rho (\mathbf{r})\rho (\mathbf{r}^{\prime })}{|%
\mathbf{r}-\mathbf{r}^{\prime }|}  \notag \\
&&+E_{xc}^{LSDA}[\rho ,\mathbf{m}]+E^{Model}[\hat{n}_{ab}]-E_{DC}^{Model}[%
\hat{n}_{ab}].  \label{SDFtot}
\end{eqnarray}%
The energies $\epsilon _{\mathbf{k}j}$ are determined by Pauli--like
Kohn--Sham matrix equation: 
\begin{equation}
\lbrack (-\nabla ^{2}+V_{eff})\hat{I}+g\mu _{B}\mathbf{B}_{eff}\hat{\mathbf{s%
}}+\xi \mathbf{l}\hat{\mathbf{s}}+\sum_{ab}\Delta \hat{V}_{ab}|\phi
_{a}\rangle \langle \phi _{b}|]\,\vec{\psi}_{\mathbf{k}j}=\epsilon _{\mathbf{%
k}j}\,\vec{\psi}_{\mathbf{k}j}  \label{SDFpau}
\end{equation}%
Here $\hat{\mathbf{l}}$ and $\hat{\mathbf{s}}$ are one--electron orbital and
spin angular momentum operator, respectively. $\xi (\mathbf{r})$ determines
the strength of spin--orbit coupling and in practice is determined~\cite%
{harmon} by radial derivative of the $l=0$ component of the Kohn--Sham
potential inside an atomic sphere: 
\begin{equation}
\xi (r)=\frac{2}{c^{2}}\frac{dV_{eff}(r)}{dr}.  \label{SDFsoc}
\end{equation}%
$E_{xc}^{LSDA}[\rho ,\mathbf{m}]$ is the LSDA exchange correlation energy.
When magnetization is present, the exchange--correlation energy functional
is assumed to be dependent on density and absolute value of the
magnetization: 
\begin{eqnarray}
&&E_{xc}^{LSDA}[\rho ,\mathbf{m}]=\int d\mathbf{r}\epsilon _{xc}[\rho (%
\mathbf{r}),m(\mathbf{r})]\rho (\mathbf{r})  \notag \\
&&+\int d\mathbf{r}f_{xc}[\rho (\mathbf{r}),m(\mathbf{r})]{\ m}(\mathbf{r}),
\label{SDFlsd}
\end{eqnarray}%
where $m(\mathbf{r})=|\mathbf{m}(\mathbf{r})|$.

$E^{Model}[\hat{n}_{ab}]$ is a contribution from the Coulomb energy in the
shell of correlated electrons 
\begin{eqnarray}
&&E^{Model}[\hat{n}_{ab}]=\frac{1}{2}\sum_{\sigma }\sum_{abcd}\langle
ac|v_{C}|bd\rangle n_{ab}^{\sigma \sigma }n_{cd}^{-\sigma -\sigma }  \notag
\\
&&+\frac{1}{2}\sum_{\sigma }\sum_{abcd}(\langle ac|v_{C}|bd\rangle -\langle
ac|v_{C}|db\rangle )n_{ab}^{\sigma \sigma }n_{cd}^{\sigma \sigma }  \notag \\
&&-\frac{1}{2}\sum_{\sigma }\sum_{abcd}\langle ac|v_{C}|db\rangle
n_{ab}^{\sigma -\sigma }n_{cd}^{-\sigma \sigma }  \label{SDFmod}
\end{eqnarray}%
This expression is nothing else as a Hartree--Fock average of the original
expression for the Coulomb interaction 
\begin{equation}
\frac{1}{2}\sum_{\sigma \sigma ^{\prime }}\sum_{abcd}\langle
ab|v_{C}|cd\rangle c_{a\sigma }^{\dagger }c_{b\sigma ^{\prime }}^{\dagger
}c_{d\sigma ^{\prime }}c_{c\sigma }  \label{SDFint}
\end{equation}%
where the Coulomb interaction $v_{C}(\mathbf{r}-\mathbf{r}^{\prime })$ has
to take into account the effects of screening by conduction electrons. The
matrix $\langle ac|v_{C}|bd\rangle =U_{abcd}$ is standardly expressed via
the Slater integrals which for d--electrons are three constants $%
F^{(0)},F^{(2)},$ and $F^{(4)}$ considered as the external parameters of the
method. Its determination can for example be done using atomic spectral
data, constrained density functional theory calculations or taken from
spectroscopic experiments.

Since part of this Coulomb energy is already taken into account in LDA
functional, the double--counting part denoted $E_{DC}^{Model}[\hat{n}_{ab}]$
must be subtracted. Frequently used form~\cite{LDA+U} for $E_{DC}^{Model}[%
\hat{n}_{ab}]$ is 
\begin{equation}
E_{DC}^{Model}[\hat{n}_{ab}]=\frac{1}{2}\bar{U}\bar{n}(\bar{n}-1)-\frac{1}{2}%
\bar{J}[\bar{n}^{\uparrow }(\bar{n}^{\uparrow }-1)+\bar{n}^{\downarrow }(%
\bar{n}^{\downarrow }-1)]  \label{SDFedc}
\end{equation}%
where 
\begin{eqnarray}
&&\bar{U}=\frac{1}{(2l+1)^{2}}\sum_{ab}\langle ab|\frac{1}{r}|ab\rangle
\label{SDFuav} \\
&&\bar{J}=\bar{U}-\frac{1}{2l(2l+1)}\sum_{ab}(\langle ab|\frac{1}{r}%
|ab\rangle -\langle ab|\frac{1}{r}|ba\rangle )  \label{SDFjav}
\end{eqnarray}%
and where $\bar{n}^{\sigma }=\sum_{a}n_{aa}^{\sigma \sigma }$ and $\bar{n}=%
\bar{n}^{\uparrow }+\bar{n}^{\downarrow }$.

The Kohn--Sham potential $V_{eff}(\mathbf{r})$ and Kohn--Sham magnetic field 
$\mathbf{B}_{eff}(\mathbf{r})$ are obtained by extremizing the functional
with respect to $\rho (\mathbf{r})$ and $\mathbf{m}(\mathbf{r})$: 
\begin{equation}
V_{eff}(\mathbf{r})=V_{ext}(\mathbf{r})+\int d\mathbf{r}^{\prime }\frac{\rho
(\mathbf{r}^{\prime })}{\left\vert \mathbf{r}-\mathbf{r}^{\prime
}\right\vert }+\frac{\delta E_{xc}^{LSDA}[\rho ,\mathbf{m}]}{\delta \rho (%
\mathbf{r})},  \label{SDFvks}
\end{equation}%
\begin{equation}
\mathbf{B}_{eff}(\mathbf{r})=\mathbf{B}_{ext}(\mathbf{r})+\frac{\delta
E_{xc}^{LSDA}[\rho ,\mathbf{m}]}{\delta \mathbf{m}(\mathbf{r})}.
\label{SDFbks}
\end{equation}%
Extremizing with respect to $n_{ab}^{\sigma \sigma ^{\prime }}$ yield the
correction to the potential $\Delta V_{ab}^{\sigma \sigma ^{\prime }}$ 
\begin{equation}
\Delta V_{ab}^{\sigma \sigma ^{\prime }}=\frac{dE^{Model}}{dn_{ab}^{\sigma
\sigma ^{\prime }}}-\frac{dE_{DC}^{Model}}{dn_{ab}^{\sigma \sigma ^{\prime }}%
}  \label{SDFlam}
\end{equation}%
Then the spin diagonal elements for potential correction are given by: 
\begin{eqnarray}
\Delta V_{ab}^{\sigma \sigma } &=&\sum_{cd}U_{abcd}n_{cd}^{-\sigma -\sigma
}+\sum_{cd}(U_{abcd}-U_{adcb})n_{cd}^{\sigma \sigma }  \notag \\
&&-\delta _{ab}\bar{U}(\bar{n}-\frac{1}{2})+\delta _{ab}\bar{J}(\bar{n}%
^{\sigma \sigma }-\frac{1}{2}),  \label{SDFlss}
\end{eqnarray}%
and spin off--diagonal elements are given by 
\begin{equation}
\Delta V_{ab}^{\sigma -\sigma }=-\sum_{cd}U_{adcb}n_{cd}^{-\sigma \sigma }
\label{SDFlsm}
\end{equation}%
The off--diagonal elements of the potential correction only present when
spin--orbit coupling is included, hence a relativistic effect.

This completes the description of the method and we now turn out to several
applications of it.

\section{Magnetic Anisotropy of Ferromagnets}

One of the oldest problems which has been addressed in the past is the
calculation of the magneto--crystalline anisotropy energy (MAE)~\cite%
{vanVleck:37,brooks,fletcher,sloncewskij,asdente} of magnetic materials. The
MAE is defined as the difference of total energies with the orientations of
magnetization pointing in different, e.g., (001) and (111), crystalline
axis. The difference is not zero because of spin--orbit effect, which
couples the magnetization to the lattice, and determines the direction of
magnetization, called the easy axis.

Being a ground state property, the MAE should be accessible in principle via
spin density functional theory described above, and, despite the primary
difficulty related to the smallness of MAE ($\sim 1\ \mu eV/$atom), great
efforts to compute the quantity with advanced total energy methods combined
with the development of faster computers, have seen success in predicting
its correct orders of magnitudes~ \cite{efn,halilov,jansen,tjew,dks}.
However, the correct easy axis of Ni has not been predicted by the LSDA and
a great amount of work has been done to understand what is the difficulty.
These include (i) scaling spin--orbit coupling in order to enlarge its
effect on the MAE~\cite{halilov,jansen}, (ii) calculating torque to avoid
comparing large numbers of energy~\cite{jansen}, (iii) studying the effects
of the second Hund's rule in the orbital polarization theory ~\cite{tjew},
(iv) analyzing possible changes in the position of the Fermi level by
changing the number of valence electrons ~\cite{dks}, (v) using the state
tracking method~\cite{freeman}, and (iv) real space approach~\cite{beiden}.

It was recently suggested that the deficiency of the calculations is lying
in improper treatment of the correlation effects and the intra--atomic
repulsion $U$ and exchange $J$ should be taken into account. It is important
to perform the calculations for fixed values of magnetic moments which
themselves show some dependency on $U$ and $J$ as studied previously \cite%
{Kudrnovsky}. Since the pure LSDA result ($U$=0, and $J$=0) reproduces the
experimental values for magnetic moments in both Fe and Ni fairly well, the $%
U-J$ parameter space should be scanned and the path of $U$ and $J$ values
which hold the theoretical moments close to the experiment was extracted.

\begin{figure}[tbh]
\epsfig{file=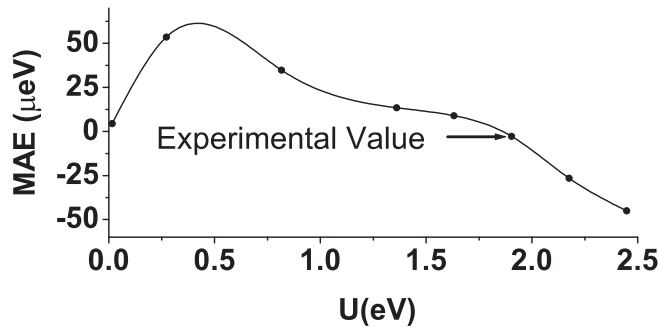,width=0.48\textwidth}
\caption{The magneto--crystalline anisotropy energy $\mbox{MAE}%
=E(111)-E(001) $ for Ni as function of $U$. The experimental MAE is marked
by arrow ($-2.8\ \protect\mu eV$). The values of exchange parameter $J$ for
every value of $U$ are chosen to hold the magnetic moment of $0.61\ \protect%
\mu _{B}$}
\label{figMAENi}
\end{figure}

The effect of correlations was found to be crucial in predicting the correct
axis of Ni. Fig.\ \ref{figMAENi} shows the results of this calculated MAE as
a function of Coulomb parameter $U$. Walking along the path of parameters $U$
and $J$ which hold the magnetic moment to 0.6 $\mu _{B}$ the MAE first
increases to $60\ \mu eV$ ($U=0.5\ eV$, $J=0.3\ eV$) and then decreases.
While decreasing it makes a rather flat area from $U=1.4\ eV$, $J=0.9\ eV$
to $U=1.7\ eV$, $J=1.1\ eV$ where MAE is positive and around $10\ \mu eV$.
After the flat area, the MAE changes from the wrong easy axis to the correct
easy axis. The correct magnetic anisotropy is predicted at $U=1.9\ eV$ and $%
J=1.2\ eV$. The change from the wrong easy axis to the correct easy axis
occurs over the range of $\delta U\sim 0.2eV$, which is of the order of
spin-orbit coupling constant ($\sim 0.1eV$).

\begin{figure}[tbh]
\epsfig{file=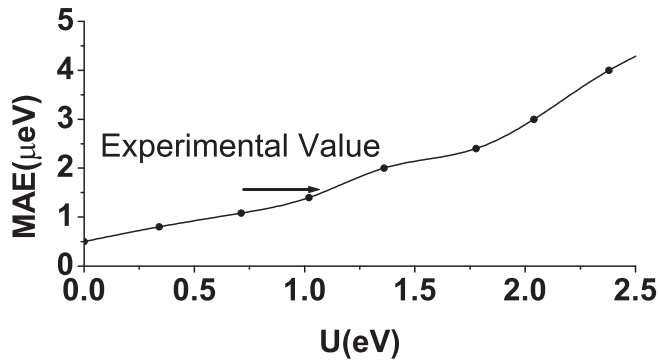,width=0.48\textwidth}
\caption{The magneto--crystalline anisotropy energy $\mbox{MAE}%
=E(111)-E(001) $ for Fe as function of $U$. The experimental MAE is marked
by arrow ($1.4\ \protect\mu eV$). The values of exchange parameter $J$ for
every value of $U$ are chosen to hold the magnetic moment of $2.2\protect\mu %
_{B}$}
\label{figMAEFe}
\end{figure}

For Fe (see Fig.\ \ref{figMAEFe}), the MAE was calculated along the path of $%
U$ and $J$ values which fixes the magnetic moment to $2.2\ \mu _{B}$. At $%
U=0\ eV$ and $J=0\ eV$, the MAE is $0.5\ \mu eV$. The correct MAE with the
correct direction of magnetic moment is predicted at $U=1.2\ eV$ and $J=0.8\
eV$. It is remarkable that the values of $U$ and $J$ necessary to reproduce
the correct magnetic anisotropy energy are very close to the values which
are needed to describe photoemission spectra of these materials~\cite{kl99}.

Recently, there has been a lot of experimental and theoretical studies
devoted to CrO$_{2}$~\cite{Sc86,Lewis97,Korotin97,Mazin99,Craco2003}. This
compound has unusual half-metallic nature, it is a metal in one spin channel
and an insulator in the other. CrO$_{2}\,\,$ demonstrates ferromagnetic
ordering with magnetic moment $2\mu _{b}$ per Cr atom and Curie Temperature $%
T_{C}\simeq 390K$. A number of published works~\cite%
{Sc86,Lewis97,Korotin97,Mazin99,Craco2003} addressed its band structure.

\begin{figure}[tbh]
\epsfig{file=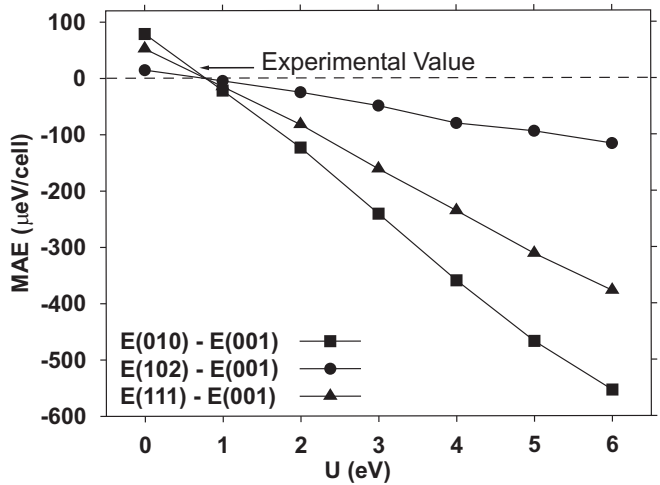,width=0.48\textwidth}
\caption{The magneto--crystalline anisotropy energies $\mbox{MAE}$ for CrO$%
_{2}\,\,$ as functions of $U$. The experimental value of MAE $%
E[010]-E[001]=15.6\protect\mu eV$ per cell is shown by arrow.}
\label{figMAECrO2}
\end{figure}

The results of the MAE computations within LSDA have been studied recently.
Total energy calculations for three different directions [001], [010] and
[102]. [001] axis indicate that easy magnetization axis within LSDA is
consistent with latest thin film experiments \cite{Sp00,Ya00,Li99}. The
numerical values of MAE however exceed the experimental one approximately
two times.

In order to figure out the influence of intra-atomic repulsion $U$ on the
magnetic anisotropy, calculations have been also performed for finite value
of $U$ changing it from $0$ to $6eV$ ($J=0.87eV$ has been kept constant).
These results are presented in Fig.~\ref{figMAECrO2}. MAE is decreasing
rapidly starting from LDA value $\approx 68\mu eV$ per cell and changes its
sign around $U\approx 0.8eV$. This leads to switching of correct easy
magnetization axis [001] to the wrong one, namely [102]. The biggest
experimental value of MAE reported in the literature is $15.6\mu eV$ per
cell~\cite{Li99}. The calculated MAE approaches this value around point $%
U=0.6eV$. To summarize, the LSDA+U approach with $U\approx 0.6eV$ and $%
J=0.87eV$ adequately describes the magneto--crystalline anisotropy of CrO$%
_{2}$.

We can conclude that the calculations of magnetic anisotropy energies in
materials are only in its beginning stage of development .It is by now clear
that the effects of Coulomb correlations must be taken seriously into
account but the smallness of MAE and its sensitivity to the values of $U$
used in the calculations prompts us that a predictive power of the method is
yet to be achieved. More work is clearly required to develop robust
algorithms calculating this highly important property of ferromagnets.

\section{Spin Spiral Method and Frozen Magnon Calculations}

While ferromagnetic spin alignments are relatively simple to explore by the
methods of spin density functional theory, the situation gets formally much
more complicated when non--collinearity occurs. The simplest example of
non--collinearity is the antiferromagnetism which is the ordering with
wave-vector $\mathbf{q}$ corresponding to a zone--boundary point of the
Brillouin zone. This requires a choice of doubled unit cell so that in the
new lattice this vector $\mathbf{q}$ becomes one of the reciprocal lattice
vectors. In a more general sense we can consider an appearance of a spin
spiral which gives orientation of different magnetic moments $\mathbf{M}%
_{\tau +R}$ for atoms of the sublattice $\mathbf{\tau }+\mathbf{R}$ ($%
\mathbf{\tau }$ is the basis vector, $\mathbf{R}$ is the primitive
translation) in the form%
\begin{equation}
\mathbf{M}_{\tau +R}=\left\{ 
\begin{array}{c}
M_{\tau +R}^{x} \\ 
M_{\tau +R}^{y} \\ 
M_{\tau +R}^{z}%
\end{array}%
\right\} =M_{\tau }\left\{ 
\begin{array}{c}
\cos (\mathbf{qR}+\phi _{\tau })\sin (\theta _{\tau }) \\ 
\sin (\mathbf{qR}+\phi _{\tau })\sin (\theta _{\tau }) \\ 
\cos (\theta _{\tau })%
\end{array}%
\right\} ,  \label{SSMmag}
\end{equation}%
where $M_{\tau }$ is the length of the atomic moment, and $\theta _{\tau
},\phi _{\tau }$ are the angles describing its orientation in the unit cell
with $\mathbf{R}=0.$ As we see, in the systems with non--collinear ordering
the situation gets immediately very complicated as \ for arbitrary $\mathbf{q%
}$ the size of the unit cell becomes prohibitively large. Note that here we
do not assume that the size of the moment is varied when going from one cell
to another. The latter is another type of so called incommensurate
magnetism, the most known example of which is Cr.

Here we will briefly review a very elegant way to solve the problem of
non--collinear magnetism where the size of the moment is kept constant which
is known as the spin spiral method developed by Sandratski \cite{PhysRepSan}%
. The problem when the size itself varies is more difficult one and the use
of linear response theory will be discussed later of this review. The formal
point is to note that non--collinear magnetic state destroys the periodicity
of the original lattice and therefore the original Bloch theorem applied to
the Kohn--Sham states $\vec{\psi}_{\mathbf{k}j}(\mathbf{r})$ is no longer
available. However, the Bloch theorem in the present form reflects the
translational symmetry properties of the lattice only it does not take into
account the symmetry properties of the spin subsystem. A generalized
symmetry treatment of the Hamiltonian for the non--collinear magnet can be
developed. Let us define group operations acting on a spinor state $\vec{\psi%
}(\mathbf{r}).$ These are (i) the translational operator $\hat{T}_{R}$ so
that $\hat{T}_{R}\mathbf{r}=\mathbf{r}+\mathbf{R,}$ (ii) the generalized
rotation $\hat{g}$ which is a pure rotation $\hat{a}=(\alpha _{g},\beta
_{g},\gamma _{g})$ described by the 3x3 $j=1$ Wigner matrix $U_{mm^{\prime
}}^{j=1}(\alpha _{g},\beta _{g},\gamma _{g})$ (where $\alpha _{g},\beta
_{g},\gamma _{g}$ are the Euler angles) and possible shift by vector $%
\mathbf{b}$ for non--symmorphic group so that $\hat{g}\mathbf{r=}\hat{a}%
\mathbf{r+b,}$ (iii) the rotation $\hat{\xi}=(\alpha _{\xi },\beta _{\xi
},\gamma _{\xi })$ in the spin--$\frac{1}{2}$ subspace described by the 2x2 $%
j=\frac{1}{2}$ Wigner matrix $U_{\sigma \sigma ^{\prime }}^{j=1/2}(\alpha
_{\xi },\beta _{\xi },\gamma _{\xi })$.

The usefulness of these operations becomes immediately transparent if we
explore symmetry properties of non--relativistic Hamiltonians $H_{\sigma
\sigma ^{\prime }}(\mathbf{r})=(-\nabla ^{2}+V_{eff})\delta _{\sigma \sigma
^{\prime }}+g\mu _{B}\mathbf{B}_{eff}\mathbf{s}_{\sigma \sigma ^{\prime }}$
of the type (\ref{SDFpau}) with non--collinear spin alignments which assumes
that the effective magnetic field $\mathbf{B}_{eff}$ inside each atom
centered at $\mathbf{\tau }+\mathbf{R}$ transforms similar to (\ref{SSMmag}%
). The translations $\mathbf{R}$ combined with the spin rotations $\hat{\xi}%
_{\{qR\tau \}}=(\theta _{\tau },\mathbf{qR}+\phi _{\tau },0)$ described by
the unitary matrices $\hat{U}^{j=1/2}$ leave the spin spiral structure
invariant. We thus arrive to a generalized Bloch theorem:%
\begin{equation}
\vec{\psi}_{\mathbf{k}j}(\mathbf{r+R})=e^{i\mathbf{kR}}\hat{U}%
^{j=1/2}(\theta _{\tau },\mathbf{qR}+\phi _{\tau },0)\vec{\psi}_{\mathbf{k}%
j}(\mathbf{r})  \label{SSMblo}
\end{equation}

Remarkably that this property allows us to restrict the consideration to a
chemical unit cell and not to the supercell. Hence, instead of introducing
the supercell whose size depends on the symmetry of wave vector $\mathbf{q}$%
, the non--collinear spin spiral states can still be treated with the
Hamiltonians whose size does not depend on the wavevector $\mathbf{q}.$

\begin{figure}[tbh]
\epsfig{file=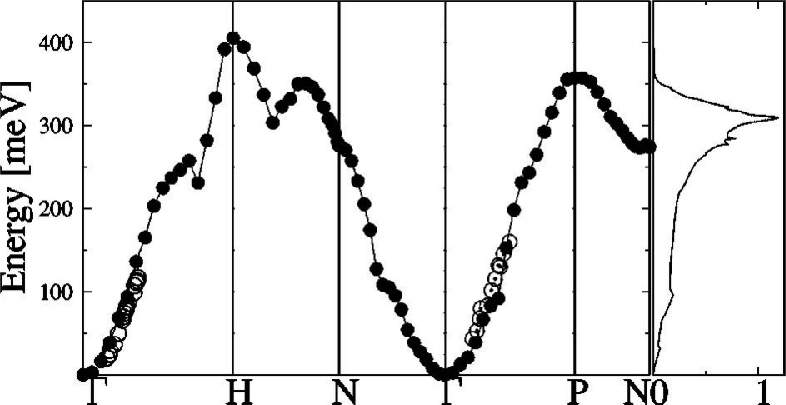,width=0.48\textwidth}
\caption{Comparison between calculated using frozen magnon method and
experimental \protect\cite{ExpFe} spin wave spectrum for Fe.}
\label{figSWFe}
\end{figure}

A variety of different calculations has been done with the spin spiral
approach \cite{PhysRepSan,JMMAntropov}. Halilov at. al \cite%
{Halilov,HalilovPRB} have studied adiabatic spin waves computed using a
frozen magnon method. This is analogous to the frozen phonon method where
atoms are displaced from their equilibrium positions and the total energy is
restored as a function of these positions. Here the calculation is
simplified since there is no necessity to introduce supercells, and can be
performed for any general spin wave-vector $\mathbf{q}$. Fig. \ref{figSWFe}
reproduces the comparison between calculated using frozen magnon LSDA
approach and experimental \cite{ExpFe} spin wave spectra for Fe.

\begin{figure}[tbh]
\epsfig{file=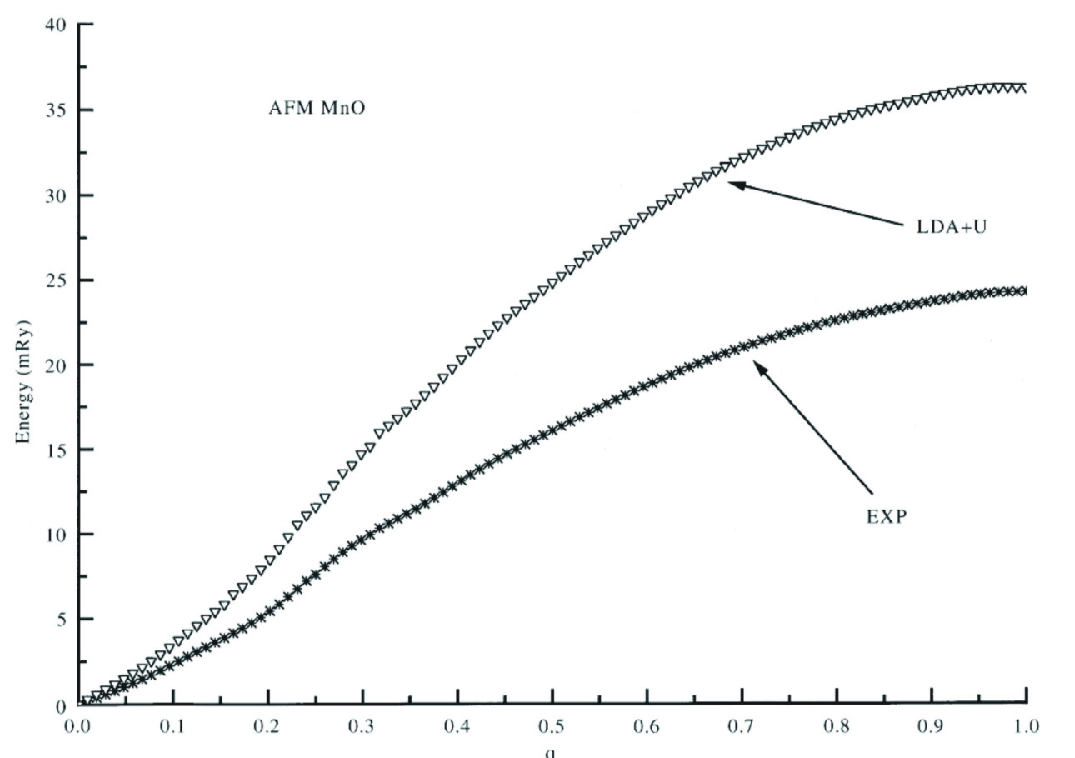,width=0.48\textwidth}
\caption{Comparison between calculated using LDA+U method and experimental
spin wave spectrum for NiO.}
\label{figSWNiO}
\end{figure}
As we see the obtained agreement in these calculations is very satisfactory
despite the fact that LSDA\ theory does not take into account important
effects of dynamical correlations. For systems like NiO, the correlation
effects become more important and the frozen magnon calculations using
LSDA+U method have been carried out. Fig. \ref{figSWNiO} shows the
comparison between calculated using LSDA+U\ method and experimental spin
wave spectra for NiO. The overall agreement is good but important
discrepancies remain which point out on the necessity to treat correlations
effects among d--electrons beyond its simplest Hartree--Fock treatment as it
is done within LSDA+U.

\section{Calculations of Exchange Constants}

While the density functional theory provide a rigorous description of the
ground state properties of magnetic materials, the finite--temperature
magnetism is estimated following a simple suggestion \cite{Licht}, whereby
constrained DFT at $T=0$ is used to extract exchange constants for a \textit{%
classical} Heisenberg model, which in turn is solved using approximate
methods (e.g. RPA, mean field) from a classical statistical mechanics of
spin systems \cite{Licht,Rosengaard, Halilov, AntropovPRL}. The recent
implementation of this approach gives reasonable value of transition
temperature for iron but not for nickel \cite{Pajda}. The analysis of
exchange parameters for different classes of magnetic materials such as
dilute magnetic semiconductors \cite{DMS}, molecular magnets \cite{molmag},
colossal magnetoresistance perovskites \cite{CMR}, transition metal alloys 
\cite{alloy}, hard magnetic materials such as PtCo \cite{PtCo} gives a
useful information for magnetic simulations.

In this section we outline the general strategy for estimations of exchange
interactions in solids. The most reliable way to consider spin excitations
of itinerant electron magnets in the framework of the spin density
functional approach is the use of frequency dependent magnetic
susceptibility within the linear response theory \cite%
{callaway,CookeNiFe,SPINPRL}.

\begin{equation}
\delta \mathbf{m}=\hat{\chi}\delta \mathbf{B}_{ext}=\hat{\chi}_{0}\delta 
\mathbf{B}_{eff},  \label{EXCchi}
\end{equation}%
where $\chi _{0}$ is a \textquotedblleft bare\textquotedblright\ DFT and $%
\chi $ is an enhanced susceptibility. In collinear magnetic structures there
are no coupling between the longitudinal and transverse components and for
the transverse spin susceptibility we have the following equation: 
\begin{equation}
\chi ^{+-}(\mathbf{r,r}^{\prime },\omega )=\chi _{0}^{+-}(\mathbf{r,r}%
^{\prime },\omega )+\int d\mathbf{r}^{\prime \prime }\chi _{0}^{+-}(\mathbf{%
r,r}^{\prime \prime },\omega )I(\mathbf{r}^{\prime \prime })\chi ^{+-}(%
\mathbf{r}^{\prime \prime }\mathbf{,r}^{\prime },\omega )  \label{transverse}
\end{equation}%
where $I=B_{xc}/m$ is an exchange-correlation \textquotedblleft Hund's
rule\textquotedblright\ interaction. This random--phase--approximation (RPA)
like equation is formally exact in the adiabatic time dependent version of
density functional theory (TD--DFT) \cite{TD-DFT}. The bare susceptibility
has the following form: 
\begin{equation}
\chi _{0}^{+-}(\mathbf{r,r}^{\prime },\omega )=\sum_{\mu \nu }\frac{f_{\mu
\uparrow }-f_{\nu \downarrow }}{\omega -\varepsilon _{\mu \uparrow
}+\varepsilon _{\nu \downarrow }}\psi _{\mu \uparrow }^{\ast }(\mathbf{r)}%
\psi _{\nu \downarrow }(\mathbf{r)}\psi _{\nu \downarrow }^{\ast }(\mathbf{r}%
^{\prime }\mathbf{)}\psi _{\mu \uparrow }(\mathbf{r}^{\prime }\mathbf{)}
\label{empty}
\end{equation}%
where $\psi _{\mu \sigma }$ and $\varepsilon _{\mu \sigma }$ are eigenstates
and eigenvalues for the Kohn--Sham quasiparticles. We can rewrite the
equation for the transverse susceptibility as 
\begin{equation}
\widehat{\chi }^{+-}=\left( m+\widehat{\Lambda }\right) \left( \omega -I_{xc}%
\widehat{\Lambda }\right) ^{-1}  \label{chi_final}
\end{equation}%
where 
\begin{equation}
\Lambda (\mathbf{r,r}^{\prime },\omega )=\sum_{\mu \nu }\frac{f_{\mu
\uparrow }-f_{\nu \downarrow }}{\omega -\varepsilon _{\mu \uparrow
}+\varepsilon _{\nu \downarrow }}\psi _{\mu \uparrow }^{\ast }(\mathbf{r)}%
\psi _{\nu \downarrow }(\mathbf{r)}\nabla \left[ \psi _{\mu \uparrow }(%
\mathbf{r}^{\prime }\mathbf{)}\nabla \psi _{\nu \downarrow }^{\ast }(\mathbf{%
r}^{\prime }\mathbf{)}-\psi _{\nu \downarrow }^{\ast }(\mathbf{r}^{\prime }%
\mathbf{)}\nabla \psi _{\mu \uparrow }(\mathbf{r}^{\prime }\mathbf{)}\right]
\label{lambda}
\end{equation}

Spin wave excitations can be separated from the Stoner continuum (e.g.,
paramagnons) only in the adiabatic approximation, which means the
replacement $\Lambda(\mathbf{r,r}^{\prime},\omega)$ by $\Lambda(\mathbf{r,r}%
^{\prime},0)$ in Eq.(\ref{chi_final}). Otherwise one should just find the
poles of the total susceptibility, and the whole concept of ``exchange
interactions''\ is not uniquely defined. Nevertheless, \textit{formally} we
can introduce the effective exchange interactions via the quantities 
\begin{equation}
\Omega(\mathbf{r,r}^{\prime},\omega)=I_{xc}\Lambda(\mathbf{r,r}^{\prime
},\omega).  \label{A1}
\end{equation}
In the static limit one can show that 
\begin{equation}
\Omega(\mathbf{r,r}^{\prime},0)=\frac{1}{m(\mathbf{r)}}J(\mathbf{r,r}%
^{\prime },0)-B_{xc}(\mathbf{r)\delta}(\mathbf{r-r}^{\prime})  \label{A2}
\end{equation}
where an expression for frequency dependent exchange interactions has the
following form 
\begin{equation}
J(\mathbf{r,r}^{\prime},\omega)=\sum_{\mu\nu}\frac{f_{\mu\uparrow}-f_{\nu%
\downarrow}}{\omega-\varepsilon_{\mu\uparrow}+\varepsilon _{\nu\downarrow}}%
\psi_{\mu\uparrow}^{\ast}(\mathbf{r)}B_{xc}(\mathbf{r)}\psi_{\nu\downarrow}(%
\mathbf{r)}\psi_{\nu\downarrow}^{\ast}(\mathbf{r}^{\prime}\mathbf{)}B_{xc}(%
\mathbf{r}^{\prime}\mathbf{)}\psi_{\mu\uparrow }(\mathbf{r}^{\prime}\mathbf{)%
}  \label{J}
\end{equation}
The later coincides with the exchange integrals \cite{Licht,JMMAntropov} if
we neglect the $\omega$ - dependence. Since $B_{xc}\sim m$ we have $J\sim
m^{2}$ and the expression (\ref{A1}) vanishes in non-magnetic case, as it
should be. Note that the static susceptibility $\widehat{\chi}^{+-}\left(
\omega =0\right) $ can be represented in the following form 
\begin{equation}
\widehat{\chi}^{+-}\left( 0\right) =m\left( \widehat{\Omega}^{-1}-{%
B_{xc}^{-1}}\right) \equiv m\widehat{\widetilde{\Omega}}  \label{static}
\end{equation}
which is equivalent to the result of Ref. \cite{Bruno} 
\begin{equation}
\widehat{\widetilde{\Omega}}=\widehat{\Omega}\left( 1-B_{xc}^{-1}\widehat{%
\Omega}\right) ^{-1}  \label{bruno}
\end{equation}
for the renormalized exchange interaction if one define them in terms of
inverse \textit{static} susceptibility \cite{antr}. Note, that the magnon
frequencies are just eigenstates of the operator $\widehat{\Omega}\left(
0\right) $ which exactly corresponds to the expression from the
unrenormalized exchange interactions \cite{Licht,JMMAntropov}. Note that for
the long-wavelength limit $\mathbf{q}\rightarrow0$ this result turns out to
be exact which proves the above statement about the stiffness constant $D$:
in the framework of the local approximation it is rigorous.

Let us compare the Eq.\ref{bruno} with the Heisenberg--like dispersion law
in the spherical approximation for the spin-wave spectrum in terms of static
transverse susceptibility\cite{JMMAntropov}: 
\begin{equation}
\omega_{\mathbf{q}}=m\left[ \chi_{\mathbf{q}}^{-1}-\chi_{0}^{-1}\right] .
\label{EXChei}
\end{equation}
Now we will use an expression $\widehat{J}=\widehat{\chi}^{-1}$which gives
the well--known connection between the inverse static susceptibility and the
parameter $\Phi_{2}$ in the Landau theory of phase transitions. To analyze
the relationship between the dispersion law (\ref{EXChei}) and that commonly
used in the DFT, $\omega_{\mathbf{q}}^{l}=mI\left[ \chi_{0}-\chi_{\mathbf{q}}%
\right] I$, we assume that the ratio%
\begin{equation}
\Delta_{\mathbf{q}}=\left( \chi_{\mathbf{q}}{}-\chi_{0}\right)
\chi_{0}^{-1}\approx\omega_{\mathbf{q}}^{l}/mI  \label{EXCdel}
\end{equation}
is small \ in the long-wave approximation. Then, by expanding Eq.\ref{EXChei}
over the parameter $\Delta_{\mathbf{q}},$ we obtain the desired result:%
\begin{align}
\omega_{\mathbf{q}} & =m\left( \chi_{\mathbf{q}}^{-1}-\chi_{0}^{-1}\right)
=m\chi_{0}^{-1}\Delta_{\mathbf{q}}\left( 1-\Delta_{\mathbf{q}}\right) ^{-1} 
\notag \\
& \simeq m\chi_{0}^{-1}\left( \chi{}_{0}-\chi_{\mathbf{q}}\right) \chi
_{0}^{-1}=m\left( J_{0}^{l}-J_{\mathbf{q}}^{l}{}\right)  \label{EXComq}
\end{align}
or%
\begin{equation}
J_{\mathbf{q}}=J_{0}\left[ 1+\Delta_{\mathbf{q}}\left( 1-\Delta_{\mathbf{q}%
}\right) ^{-1}\right] =J_{0}\left[ 1+\Delta_{\mathbf{q}}+\Delta _{\mathbf{q}%
}^{2}+..\right] ,  \label{EXCexp}
\end{equation}
where the susceptibility has a matrix structure and%
\begin{equation}
J_{\mathbf{q}}^{l}=J_{0}\left( 1+\Delta_{\mathbf{q}}\right)
=\chi_{0}^{-1}\chi_{\mathbf{q}}\chi_{0}^{-1}=I\chi_{\mathbf{q}}I
\label{EXCjey}
\end{equation}
is the matrix of the exchange parameter in the local (long--wave)
approximation. We take into account, that in the ferromagnetic state $\chi
_{0}^{-1}=I.$ A form of Eq.(\ref{EXCjey}) is the widely accepted definition
of the exchange coupling parameter. The usual assumption of weak enhancement 
$J_{\mathbf{q}}-J_{0}=J_{\mathbf{q}}^{0}-J_{0}^{0}$ is valid due to the
well--known result $\chi_{\mathbf{q}}^{-1}=J_{\mathbf{q}}=J_{\mathbf{q}%
}^{0}-I$. As a consequence, the spectrum of elementary excitations is not
affected by exchange--correlation enhancement effects in linear response
regime. So, the definition (\ref{EXCjey}) is directly related to several
very strong approximations: rigid spin, smallness of spin wave dispersion
compared to the effective exchange splitting (atomic limit).

Below we will briefly mention calculations of the exchange parameters and
adiabatic spin wave spectra using the multiple scattering technique in the
local density approximation with a linear muffin--tin orbital technique
(LMTO) in the atomic sphere approximation \cite{OKA}. Three ferromagnetic
systems with entirely different degree of localization of the local moment
were considered: Gd, Ni and Fe \cite{JMMAntropov}. An important observation
is that in Gd (highly localized moments, small spin wave dispersion) the
approximation based on the assumption $\Delta _{\mathbf{q}}<<1$ is
completely fulfilled ($\Delta $ at the zone$\ $boundary is less then $0.01$%
), whereas in Ni (moderately localized moments, large spin wave dispersion)
it is not valid at all ($\Delta \simeq 0.6$ at $\mathbf{q=}Y$), so that the
corresponding matrix estimates using Eq.(\ref{EXComq}) demonstrate a strong
enhancement of spin wave spectra at larger $\mathbf{q}$. Fe is an
intermediate case where the maximum of $\Delta $ is 0.30.

Practical expression for the exchange coupling has been derived \cite{antr}
using multiple scattering method: 
\begin{equation}
J_{\mathbf{q}}^{+-}=\frac{1}{4\pi }\int^{\varepsilon _{F}}d\varepsilon Im%
\mathrm{Tr}_{L}\left( T^{\uparrow }-T^{\downarrow }\right) _{00}\left[ \int d%
\mathbf{k}T_{\mathbf{k}}^{\uparrow }T_{\mathbf{k}+\mathbf{q}}^{\downarrow }%
\right] ^{-1}\left( T^{\uparrow }-T^{\downarrow }\right) _{00}.  \label{e1}
\end{equation}%
where $T^{\sigma }$ is a path operator. This expression properly takes into
account both exchange effects related to the Stoner splitting and the
different energy dispersion for different spins in band magnets.

Analysis of the effective exchange coupling between atoms after Eq. (\ref{e1}%
) indicated that in Fe and Ni the main contribution came from the
renormalization of the first nearest neighbor exchange, so that in bcc Fe it
is increased from $J_{01}^{l}$=16.6 meV to $J_{01}$=19.4 meV, whereas in fcc
Ni $J_{01}^{l}$=2.7 meV to $J_{01}$=8.3 meV. Such a clear difference in the
results indicates that the removal of the long--wave approximation can serve
as an indicator of the degree of localization (parameter $\left( \chi _{%
\mathbf{q}}{}-\chi _{0}\right) \chi _{0}^{-1}$ above) in different metallic
magnets. In Fe and Ni, for instance, it at least partially explains why long
wavelength mean field estimates predict such a small $T_{c}$ in FM Ni:
300--350K in Ref.\cite{JMMAntropov}, 340 K in present calculations, while
experimental result is 630 K. First of all, long--wave approximation is
suitable for such a 'localized' system as Fe, and the corresponding change
in the nearest neighbor $J_{01}$ is relatively small (correspondingly the
increase of $T_{c}$ is expected to be small). Ferromagnetic Ni represents a
rather itinerant system and any local approach (long wave approximation in
particular) might produce a large error. Also in Ni at $T_{c}$ the energy
associated with 'creation' of moment is comparable with the energy of its
rotation and longitudinal fluctuations should be taken into account.

A large increase in $J_{01}$ for Ni may indicate that traditional
mean--field approaches (based on the absence of short--range order) are not
applicable for the itinerant systems in general, predicting very high $T_{c}$
(above 1000K). Such a number is not consistent with the local density
approach because the latter does not allow to have $T_{c}$ in fcc Ni larger
then 500-540K (600-640K using gradient corrections) and this number can be
considered only as an indicator of non--applicability of mean field approach.

\section{Dynamical Spin Susceptibility Calculations}

Local spin density functional theory and the LDA+U method have been applied
with success to study full wave--vector and frequency dependent spin
susceptibility $\chi $ of solids$.$ Its knowledge accessible directly via
neutron--scattering measurements is important due to significant influence
of spin fluctuations to many physical properties and phenomena \cite{Moria},
such, \textit{e.g.}, as the electronic specific heat, electrical and thermal
resistivity, suppression of superconductivity for singlet spin pairing, 
\textit{etc }. As we already discussed, in magnetically ordered materials,
transverse spin fluctuations are spin waves whose energies and lifetimes are
seen in the structure of transverse susceptibility.

Large efforts in the past were put on the development of methods for \textit{%
ab initio }calculations of the dynamical spin susceptibility based either on
the random--phase--approximation (RPA)\ decoupling of the Bethe--Salpeter
equation \cite{Cooke1973}, or within density functional formalism \cite%
{Callaway1975}. Quantitative estimates of $\chi $ with realistic energy
bands, wave functions, and self--consistently screened electron--electron
matrix elements appeared in the literature more than 3 decades ago \cite%
{CookeNiFe,CallawayNi,WinterPdV}. More recently, a computationally efficient
time--dependent generalization of the Sternheimer approach \cite{LR}
originally developed for \textit{ab initio} calculations of phonon
dispersions, electron--phonon interactions and transport properties of
transition--metal materials has been proposed \cite{SPINPRL}.

This method is advantageous since it allows us to avoid problems connected
to the summations over higher energy states and inversions of large
matrices. It considers the response of electrons to a small external
magnetic field 
\begin{equation}
\delta \mathbf{B}_{ext}(\mathbf{r}t)=\delta \mathbf{b}e^{i(\mathbf{q}+%
\mathbf{G})\mathbf{r}}e^{i\omega t}e^{-\eta |t|}+c.c  \label{CHIf1}
\end{equation}%
applied to a solid. Here $\delta \mathbf{b}=\sum_{\mu }\delta b^{\mu }%
\mathbf{e}_{\mu }$ shows a polarization of the field ($\mu $ runs over $%
x,y,z\,$\ or over $-1,0,1$), wave vector $\mathbf{q}$ lies in the first
Brillouin zone, $\mathbf{G}$ is a reciprocal lattice vector, and $\eta $ is
an infinitesimal positive quantity. If the unperturbed system is described
by charge density $\rho (\mathbf{r})$ and by magnetization $\mathbf{m}(%
\mathbf{r})$, the main problem is to find self--consistently first--order
changes $\delta \rho (\mathbf{r}t)$ and $\delta \mathbf{m}(\mathbf{r}%
t)=\sum_{\nu }\delta m_{\nu }(\mathbf{r}t)\mathbf{e}^{\nu }$ induced by the
field $\delta \mathbf{B}_{ext}(\mathbf{r}t)$. If the polarization $\delta 
\mathbf{b}$ in (\ref{CHIf1}) is fixed to a particular $\mu $th direction,
and $\delta \mathbf{m}(\mathbf{r}t)$ is calculated afterwards, a $\mu $th
column of the spin susceptibility matrix $\chi _{\nu \mu }(\mathbf{r},%
\mathbf{q}+\mathbf{G},\omega )$ will be found.

Employment of \emph{time--dependent }(TD)\ version of density functional
theory (DFT) \cite{TD-DFT} to find the quantities $\delta \rho (\mathbf{r}t)$
and $\delta \mathbf{m}(\mathbf{r}t)$ can be useful because in order to find
the dynamical response within TD DFT, only the knowledge of these
unperturbed Kohn--Sham states (both occupied and unoccupied) is required; no
knowledge of real excitation spectra (both energies and lifetimes) is
necessary. This is the main advantage of such approach. Unfortunately,
within TD DFT, an accurate approximation to the kernel $I_{xc}(\mathbf{r},%
\mathbf{r}^{\prime },\omega )$ describing dynamical exchange--correlation
effects is unknown while some progress is currently been made \cite{Progress}%
. In the past, the so called adiabatic local density approximation (ALDA) 
\cite{TD-DFT} and a generalized gradient approximation \cite{GGA} (GGA) are
adopted to treat $I_{xc}(\mathbf{r},\mathbf{r}^{\prime },\omega )$. The
addition of the local U corrections has been addressed in the work \cite%
{CACUO2}.

One can develop a \textit{variational} linear--response formulation. It has
already been known that \emph{static} charge and spin susceptibilities
appeared as second--order changes in the total energy due to applied
external fields can be calculated in a variational way. This was
demonstrated long time ago \cite{Vosko} on the example of magnetic response,
and, recently \cite{LR,Gonze}, in the problem of lattice dynamics which is
an example of charge response. The proof is directly related to a powerful
\textquotedblright $2n+1$\textquotedblright\ theorem of perturbation theory
and stationarity property for the total energy itself \cite{2n+1}. Any ($%
2n+1)$th change in the total energy $E_{tot}$ involves finding only ($n)$th
order changes in one--electron wave functions $\psi _{i}$, and corresponding
changes in the charge density as well as in the magnetization. Any ($2n$)th
change in $E_{tot}$ is then variational with respect to the ($n$)th--order
changes in $\psi _{i}$.

A time--dependent generalization of these results was addressed\cite{SPINPRL}%
, where the action $S$ as a functional of $\rho (\mathbf{r}t)$ and $\mathbf{m%
}(\mathbf{r}t)$ is considered within TD DFT \cite{TD-DFT,Liu}. These
functions are expressed via Kohn--Sham spinor orbitals $\vec{\psi}_{\mathbf{k%
}j}\left( \mathbf{r}t\right) $ satisfying TD Schr\"{o}dinger's equation.
Therefore, $S$ as the stationary functional of $\vec{\psi}_{\mathbf{k}%
j}\left( \mathbf{r}t\right) $ is considered in practice$.$ When the external
field is small, the perturbed wave function is represented as $\vec{\psi}_{%
\mathbf{k}j}\left( \mathbf{r}\right) e^{-i\epsilon _{\mathbf{k}j}t}+\delta 
\vec{\psi}_{i}\left( \mathbf{r}t\right) $ and the first--order changes $%
\delta \vec{\psi}_{\mathbf{k}j}\left( \mathbf{r}t\right) $ define the
induced charge density as well as the magnetization: 
\begin{eqnarray}
\delta \rho &=&\sum_{\mathbf{k}j}\left( \{\delta \vec{\psi}_{\mathbf{k}j}|I|%
\vec{\psi}_{\mathbf{k}j}\}+\{\vec{\psi}_{\mathbf{k}j}|I|\delta \vec{\psi}_{%
\mathbf{k}j}\}\right)  \label{CHIf2} \\
\delta \mathbf{m} &=&g\mu _{B}\sum_{i}\left( \{\delta \vec{\psi}_{\mathbf{k}%
j}|\mathbf{\hat{s}}|\vec{\psi}_{\mathbf{k}j}\}+\{\vec{\psi}_{\mathbf{k}j}|%
\mathbf{\hat{s}}|\delta \vec{\psi}_{\mathbf{k}j}\}\right)  \label{CHIf3}
\end{eqnarray}

In order to find $\delta \vec{\psi}_{\mathbf{k}j}\left( \mathbf{r}t\right) $%
, a time--dependent analog of the \textquotedblright $2n+1$%
\textquotedblright\ theorem is now introduced. Any ($2n+1)$th change in the
action functional $S$ involves finding only ($n)$th order changes in the TD
functions $\vec{\psi}_{\mathbf{k}j}\left( \mathbf{r}t\right) $, and
corresponding changes in charge density as well as in the magnetization. Any
($2n$)th change in $S$ is then variational with respect to the ($n$%
)th--order changes in $\vec{\psi}_{\mathbf{k}j}\left( \mathbf{r}t\right) $.
The proof is the same as for the static case \cite{2n+1} if the stationarity
property of $S$ and the standard TD perturbation theory are exploited. For
important case $n=2$, this theorem makes the second--order change $S^{(2)}$
in the action \emph{variational} with respect to the first--order changes $%
\delta \vec{\psi}_{i}\left( \mathbf{r}t\right) .$ If the perturbation has
the form (\ref{CHIf1}), $S^{(2)}$ is directly related to the real diagonal
part of the dynamical spin susceptibility $Re[\chi _{\nu \mu }(\mathbf{q}+%
\mathbf{G}^{\prime },\mathbf{q}+\mathbf{G},\omega )]_{\mathbf{G}^{\prime }=%
\mathbf{G}}$ , thus allowing its variational estimate \cite{Liu}.

The problem is now reduced to find $S^{(2)}$ as a functional of $\delta \vec{%
\psi}_{\mathbf{k}j}\left( \mathbf{r}t\right) $ and to minimize it. This will
bring an equation for $\delta \vec{\psi}_{\mathbf{k}j}\left( \mathbf{r}%
t\right) $. Any change in the action functional can be established by
straightforward varying $S$ of TD DFT \cite{TD-DFT,Liu} with respect to the
perturbation (\ref{CHIf1}). This is analogous to what is done in the static
DFT to derive, for example, the dynamical matrix \cite{LR}. $S^{(2)}$ is
found to be 
\begin{eqnarray}
&&S^{(2)}[\delta \vec{\psi}_{\mathbf{k}j}]=\sum_{\mathbf{k}j}2\left\langle
\delta \vec{\psi}_{\mathbf{k}j}\left\vert H-i\partial _{\mathbf{k}%
j}I\right\vert \delta \vec{\psi}_{\mathbf{k}j}\right\rangle +  \notag \\
&&\int \delta \rho \delta V_{eff}-\int \delta \mathbf{m}(\delta \mathbf{B}%
_{eff}+\delta \mathbf{B}_{ext})  \label{CHIf4}
\end{eqnarray}%
where the unperturbed 2$\times $2 Hamiltonian matrix $\hat{H}=(-\nabla
^{2}+V_{eff})\hat{I}-g\mu _{B}\mathbf{\hat{s}B}_{eff}+\xi \mathbf{l\hat{s}}$%
. $\delta V_{eff}$ and $\delta \mathbf{B}_{eff}$ are their first--order
changes induced by the perturbation (\ref{CHIf1}) which involve the Hartree
(for $\delta V_{eff}$) and the exchange--correlation contributions expressed
via $\delta \rho \,$and $\delta \mathbf{m}$ in the standard manner \cite%
{Callaway1975}.

The differential equation for $\delta \vec{\psi}_{\mathbf{k}j}\left( \mathbf{%
r}t\right) $ derived from the stationarity condition of\ (\ref{CHIf4}) is
given by 
\begin{equation}
(H-i\partial _{\mathbf{k}j}\hat{I})\delta \vec{\psi}_{\mathbf{k}j}+(\delta
V_{eff}\hat{I}-\mu _{B}\mathbf{\sigma }\delta \mathbf{B}_{eff})\vec{\psi}_{%
\mathbf{k}j}=0  \label{f5}
\end{equation}%
\bigskip

Applications to transverse spin fluctuations in Fe and Ni as well as
calculations of paramagnetic response in Cr and Pd have demonstrated an
efficiency of the approach. Experimental evidence of an \textquotedblright
optical\textquotedblright\ spin--wave branch for Ni \cite{ExpNi} and its
absence for Fe \cite{ExpFe} was correctly described . The dynamical
susceptibility was calculated \textit{ab initio} for paramagnetic Cr, a
highly interesting material due to its incommensurate antiferromagnetism 
\cite{Cr}. The calculation predicted a wave vector of the spin density wave
(SDW), and clarified the role of Fermi--surface nesting. Strong
long--wavelength spin fluctuations of Pd were evident from these and earlier 
\cite{WinterPdV} theoretical studies.

\begin{figure}[tbh]
\epsfig{file=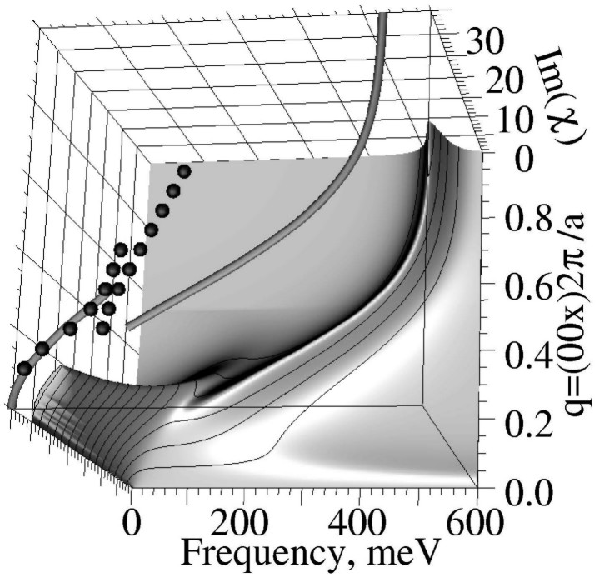,width=0.48\textwidth}
\caption{Calculated $Im[\protect\chi _{+-}(\mathbf{q},\protect\omega )]$
(arb.units) for Ni. Experimental data \protect\cite{ExpNi} are indicated by
balls.}
\label{figImXFe}
\end{figure}

We reproduce some of the calculations at Fig. \ref{figImXFe} and Fig. \ref%
{figImXCr} which show calculated $Im[\chi (\mathbf{q},\omega )]$ for
ferromagnetic Ni and paramagnetic bcc Cr. In particular, for Cr, a
remarkable structure is clearly seen for the $\mathbf{q}^{\prime }$s near
(0,0,$x_{SDW}\sim $0.9)2$\pi /a$, where the susceptibility is mostly
enhanced at low frequencies. This predicts Cr to be an incommensurate
antiferromagnet (experimentally, $x_{SDW}$=0.95).

\begin{figure}[tbh]
\epsfig{file=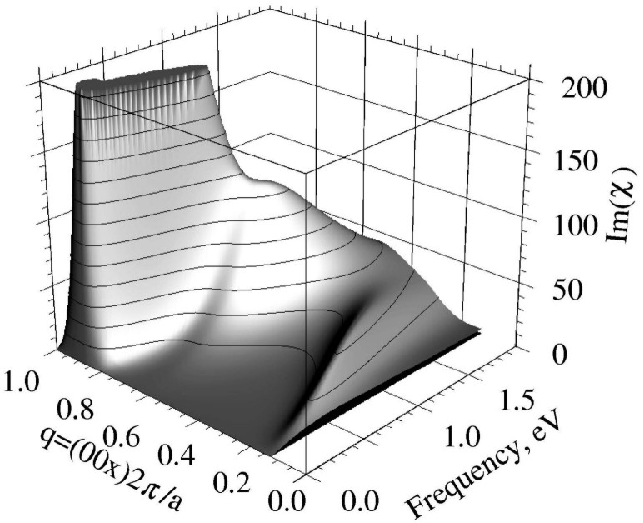,width=0.48\textwidth}
\caption{Calculated $Im[\protect\chi _{+-}(\mathbf{q},\protect\omega )]$ (Ry$%
^{-1}$) for Cr. }
\label{figImXCr}
\end{figure}

A scheme for making \textit{ab initio} calculations of the dynamic
paramagnetic spin susceptibilities of solids at finite temperatures was
recently described \cite{Staunton} where incommensurate and commensurate
antiferromagnetic spin fluctuations in paramagnetic Cr and compositionally
disordered Cr$_{95}$V$_{5}$ and Cr$_{95}$Re$_{5}$ alloys were studied
together with the connection with the nesting of their Fermi surfaces.

To conclude, the developed approach is able to describe known
spin--fluctuational spectra of some real materials which demonstrated its
efficiency for practical \textit{ab initio} calculations. however, more
elaborate approximations to the dynamical exchange and correlation are
clearly required in order to account for the observed discrepancies. The
next section discussed a more general framework to deal with the correlation
effects.

\section{Dynamical Mean Field Approach and the Regime of Strong Correlations}

Magnetic materials range from very weak having a small magnetization to
strong ones which exhibit a saturated magnetization close to the atomic
value. As we have seen in the previous sections, frequently, weak or
itinerant magnets are well described by spin density wave theory, where spin
fluctuations are localized in a small region of momentum space.
Quantitatively they are well described by LSDA. The transition from ordered
to paramagnetic state is driven by amplitude fluctuations. In strong
magnets, there is a separation of time scales, $\hbar /t$ is the time scale
for an electron to hop from site to site with hopping integral $t$, which is
much shorter than $\hbar /J,$ the time scale for the moment to flip in the
paramagnetic state. The spin fluctuations are localized in real space and
the transition to the paramagnetic state is driven by orientation
fluctuations of the spin. The exchange splitting $J$ is much larger than the
critical temperature.

Obtaining a quantitative theory of magnetic materials valid both in weak and
in strong coupling, above and below the critical temperature has been a
theoretical challenge for many years. It has been particularly difficult to
describe the regime above $T_{c}$ in strong magnets when the moments are
well formed but their orientation fluctuates. This problem, for example,
arises in magnetic insulators above their ordering temperature, when this
ordering temperature is small compared to electronic scales, a situation
that arises in transition metal monoxides (NiO and MnO) and led to the
concept of a Mott insulator. In these materials the insulating gap is much
larger than the N\'{e}el temperature. Above the ordering temperature, we
have a collection of atoms with an open shell interacting via superexchange.
This is a local moment regime which cannot be accessed easily with
traditional electronic structure methods.

Two important approaches were designed to access the disordered local moment
(DLM) regime. One approach \cite{Hubbard} starts form a Hubbard like
Hamiltonian and introduces spin fluctuations via the Hubbard--Stratonovich
transformation \cite{Stratonovich, Wang, Evenson, Cyrot} which is then
evaluated using a static coherent potential approximation (CPA) and
improvements of this technique. A dynamical CPA \cite{Al-Attar} was
developed by Kakehashi \cite{Kakehashi}. A second approach begins with
solutions of the Kohn--Sham equations of a constrained LDA approximation in
which the local moments point in random directions, and averages over their
orientation using the KKR--CPA approach \cite{Gyorffy:1979, Faulkner:1982}.
The average of the Kohn--Sham Green functions then can be taken as the first
approximation to the true Green functions, and information about angle
resolved photoemission spectra can be extracted \cite{Gyorffy:1985,
Satunton:1985}. There are approaches that are based on a picture where there
is no short range order to large degree. The opposite point of view, where
the spin fluctuations far away form the critical temperature are still
relatively long ranged was put forward in the fluctuation local band picture 
\cite{Capellmann:1974, Korenman,Prange}.

Description of the behavior near the critical point requires renormalization
group methods, and the low temperature treatment of this problem is still a
subject of intensive research \cite{Belitz:2002}. There is also a large
literature on describing magnetic metals using more standard many--body
methods \cite{Liebsch:1981,Manghi:1997,Manghi:1999,Nolting:1987,Steiner:1992}%
.

Dynamical mean field theory (DMFT)\ can be used to improve treatment to
include dynamical fluctuations beyond static approximations like LSDA or
LSDA+U. \ This is needed to get access to disordered local moment regime and
beyond. Notice that single site DMFT includes some degree of short range
correlations. Cluster methods can be used to go beyond the single site DMFT
to improve the description of short range order on the quasiparticle
spectrum. DMFT also allows us to incorporate the effects of the
electron--electron interaction on the electronic degrees of freedom. This is
relatively important in metallic systems such as Fe and Ni and absolutely
essential to obtain the Mott--Hubbard gap in transition metal monoxides.

Realistic implementation of DMFT within electronic structure framework \cite%
{Anisimov,LDA++} can be viewed as a spectral density functional theory \cite%
{Chitra1,Chitra2,Tsvelik,SDFT}. The central quantity of this method is the
local Green function $\hat{G}_{loc}(\mathbf{r},\mathbf{r}^{\prime },\omega
), $ which in the non--collinear version should be considered as the matrix
in spin variables%
\begin{equation}
\hat{G}_{loc}=\left( 
\begin{array}{cc}
G^{\uparrow \uparrow } & G^{\uparrow \downarrow } \\ 
G^{\downarrow \uparrow } & G^{\downarrow \downarrow }%
\end{array}%
\right)  \label{LOC}
\end{equation}%
The local Green function is only a part of the full one--electron Green
function restricted which is restricted within a certain cluster area. Due
to translational invariance of the Green function on the original lattice
given by primitive translations $\{\mathbf{R\}}$, i.e. $\hat{G}(\mathbf{r+R},%
\mathbf{r}^{\prime }+\mathbf{R},\omega )=\hat{G}(\mathbf{r},\mathbf{r}%
^{\prime },\omega ),$ it is always sufficient to consider $\mathbf{r}$ lying
within a primitive unit cell $\Omega _{c}$ positioned at $\mathbf{R}=0$.
Thus, $\mathbf{r}^{\prime }$ travels within some area $\Omega _{loc}$
centered at $\mathbf{R=0}$. We therefore define the local Green function to
be the exact Green function $G(\mathbf{r},\mathbf{r}^{\prime },z)$ within a
given cluster $\Omega _{loc}$ and zero outside. In other words,%
\begin{equation}
\hat{G}_{loc}(\mathbf{r},\mathbf{r}^{\prime },\omega )=\hat{G}(\mathbf{r},%
\mathbf{r}^{\prime },\omega )\theta _{loc}(\mathbf{r},\mathbf{r}^{\prime })
\label{TET}
\end{equation}%
where the theta function is a unity when vector $\mathbf{r}\in \Omega _{c},%
\mathbf{r}^{\prime }\in \Omega _{loc}$\ and zero otherwise. A functional
theory where the total free energy of the system $\Gamma _{SDF}$ considers
the local Green function as a variable was developed recently \cite%
{Chitra1,Chitra2,Tsvelik,SDFT}. It is more powerful than the ordinary DFT
since it accesses both the energetics and local excitational spectra of real
materials. The range of locality is in principle our choice and the theory
has a correct scaling in predicting full $\mathbf{k}$-- and $\omega $
dependent spectrum of excitations\ when this range is expanded till
infinity. If the range is collapsed to a single site, only the density of
states is predicted.

A numerically tractable approach was developed completely similar to the
density functional theory with the introduction of energy--dependent analog
of Kohn--Sham orbitals. They are very helpful to write down the kinetic
energy portion of the functional similar to as DFT\ considers the density
functional as the functional of Kohn--Sham wave functions. It is first
useful to introduce the notion of an local self--energy operator 
\begin{eqnarray*}
\mathcal{M}_{eff}^{\sigma \sigma ^{\prime }}(\mathbf{r,r}^{\prime },\omega )
&=&V_{ext}(\mathbf{r})\delta (\mathbf{r}-\mathbf{r}^{\prime })\delta
_{\sigma \sigma ^{\prime }}+\mathcal{M}_{int}^{\sigma \sigma ^{\prime }}(%
\mathbf{r,r}^{\prime },\omega ) \\
&=&[V_{ext}(\mathbf{r})+V_{H}(\mathbf{r})]\delta (\mathbf{r}-\mathbf{r}%
^{\prime })\delta _{\sigma \sigma ^{\prime }}+\mathcal{M}_{xc}^{\sigma
\sigma ^{\prime }}(\mathbf{r,r}^{\prime },\omega )
\end{eqnarray*}%
which has the same range of locality as the local Green function. Second, an
auxiliary Green function $\mathcal{G}^{\sigma \sigma ^{\prime }}(\mathbf{r},%
\mathbf{r}^{\prime },i\omega )$ connected to our new "interacting
Kohn--Sham" particles so that it is defined in the entire space by the
relationship 
\begin{equation*}
\lbrack \mathcal{G}^{-1}]^{\sigma \sigma ^{\prime }}(\mathbf{r},\mathbf{r}%
^{\prime },\omega )=[G_{0}^{-1}]^{\sigma \sigma ^{\prime }}(\mathbf{r},%
\mathbf{r}^{\prime },\omega )-\mathcal{M}_{int}^{\sigma \sigma ^{\prime }}(%
\mathbf{r},\mathbf{r}^{\prime },\omega ).
\end{equation*}%
where $G_{0}^{\sigma \sigma ^{\prime }}(\mathbf{r},\mathbf{r}^{\prime
},i\omega )$ is the non--interacting Green function, which is given by%
\begin{equation}
\lbrack G_{0}^{-1}]^{\sigma \sigma ^{\prime }}(\mathbf{r},\mathbf{r}^{\prime
},\omega )=[(\omega +\mu +\nabla ^{2}-V_{ext}(\mathbf{r}))\delta _{\sigma
\sigma ^{\prime }}-\xi \mathbf{l}\cdot \mathbf{s}_{\sigma \sigma ^{\prime
}}-g\mu _{B}\mathbf{B}_{ext}\mathbf{s}_{\sigma \sigma ^{\prime }}]\delta (%
\mathbf{r}-\mathbf{r}^{\prime })  \label{SDFactG11}
\end{equation}

Since $\mathcal{G}$ is a functional of $G_{loc}$, it is very useful to view
the spectral density functional $\Gamma _{SDF}$ as a functional of $\mathcal{%
G}$:%
\begin{equation}
\Gamma _{SDF}[\mathcal{G}]=\mathrm{Tr}\ln \mathcal{G}-\mathrm{Tr}[G_{0}^{-1}-%
\mathcal{G}^{-1}]\mathcal{G}+\Phi _{SDF}[G_{loc}]  \label{SDFactSD2}
\end{equation}%
where the first two contribution represent the kinetic contribution and the
contribution from the external potential and magnetic field. The unknown
interaction part of the free energy $\Phi _{SDF}[G_{loc}]$ is the functional
of $G_{loc}.$ If the Hartree term is explicitly extracted, this functional
can be represented as%
\begin{equation}
\Phi _{SDF}[G_{loc}]=E_{H}[\rho ]+\Phi _{SDF}^{xc}[G_{loc}]
\label{SDFactFXC}
\end{equation}%
\bigskip

To facilitate the calculation, the auxiliary Green function $\mathcal{G}%
^{\sigma \sigma ^{\prime }}(\mathbf{r},\mathbf{r}^{\prime },i\omega )$ can
be represented in terms of generalized energy--dependent one--particle states

\begin{equation}
\mathcal{G}^{\sigma \sigma ^{\prime }}(\mathbf{r},\mathbf{r}^{\prime
},i\omega )=\sum_{\mathbf{k}j}\frac{\psi _{\mathbf{k}j\omega }^{R(\sigma )}(%
\mathbf{r})\psi _{\mathbf{k}j\omega }^{L(\sigma ^{\prime })}(\mathbf{r}%
^{\prime })}{i\omega +\mu -E_{\mathbf{k}j\omega }}  \label{IMPeksGK1}
\end{equation}%
\bigskip which obey the one--particle Dyson equation for the "right"
eigenvector $\psi _{\mathbf{k}j\omega }^{R(\sigma )}(\mathbf{r})$ 
\begin{eqnarray}
&&\sum_{\sigma ^{\prime }}\{[-\nabla ^{2}+V_{ext}(\mathbf{r})+V_{H}(\mathbf{r%
})]\delta _{\sigma \sigma ^{\prime }}+g\mu _{B}\mathbf{B}_{eff}\mathbf{s}%
_{\sigma \sigma ^{\prime }}+\xi \mathbf{l}\cdot \mathbf{s}_{\sigma \sigma
^{\prime }}\}\psi _{\mathbf{k}j\omega }^{R(\sigma ^{\prime })}(\mathbf{r})+
\label{IMPeksDER} \\
&&\sum_{\sigma ^{\prime }}\int d\mathbf{r}^{\prime }\mathcal{M}_{xc}^{\sigma
\sigma ^{\prime }}(\mathbf{r},\mathbf{r}^{\prime },i\omega )\psi _{\mathbf{k}%
j\omega }^{R(\sigma ^{\prime })}(\mathbf{r}^{\prime })=E_{\mathbf{k}j\omega
}\psi _{\mathbf{k}j\omega }^{R(\sigma )}(\mathbf{r})  \label{IMPeksDEL}
\end{eqnarray}%
and similar equation for the "left" eigenvector $\psi _{\mathbf{k}j\omega
}^{L(\sigma )}(\mathbf{r})$ when it appears on the left.

Evaluation of the free energy requires writing down the precise functional
form for $\Phi _{SDF}[G_{loc}].$ Unfortunately, it is a problem because the
free energy $F=E-TS,$ where $E$ is the total energy and $S$ is the entropy,
and the\ evaluation of the entropy requires the evaluation of the energy as
a function of temperature and an additional integration over it. However,
the total energy formula can be written by utilizing the Galitzky--Migdal
expression for the interaction energy \cite{DMFTReview}. It is given by:

\begin{eqnarray}
&&E=T\sum_{i\omega }e^{i\omega 0^{+}}\sum_{\mathbf{k}j}g_{\mathbf{k}j\omega
}E_{\mathbf{k}j\omega }-T\sum_{i\omega }\sum_{\sigma \sigma ^{\prime }}\int d%
\mathbf{r}d\mathbf{r}^{\prime }\mathcal{M}_{eff}^{\sigma \sigma ^{\prime }}(%
\mathbf{r,r}^{\prime },i\omega )\mathcal{G}^{\sigma ^{\prime }\sigma }(%
\mathbf{r}^{\prime },\mathbf{r},i\omega )+  \notag \\
&&+\int d\mathbf{r}V_{ext}(\mathbf{r})\rho (\mathbf{r})-\int d\mathbf{rB}%
_{ext}(\mathbf{r})\mathbf{m}(\mathbf{r})+E_{H}[\rho ]+\frac{1}{2}%
T\sum_{i\omega }\sum_{\sigma \sigma ^{\prime }}\int d\mathbf{r}d\mathbf{r}%
^{\prime }\mathcal{M}_{xc}^{\sigma \sigma ^{\prime }}(\mathbf{r,r}^{\prime
},i\omega )\mathcal{G}^{\sigma ^{\prime }\sigma }(\mathbf{r}^{\prime },%
\mathbf{r},i\omega )  \label{IMPeksSDF}
\end{eqnarray}%
where%
\begin{equation}
g_{\mathbf{k}j\omega }=\frac{1}{i\omega +\mu -E_{\mathbf{k}j\omega }}
\label{IMPeksOCC}
\end{equation}%
We \ have cast the notation of spectral density theory in a form similar to
DFT. The function $g_{\mathbf{k}j\omega }$ is the Green function in the
orthogonal left/right representation which plays a role of a "frequency
dependent occupation number".

Both the charge density and non--collinear magnetization can be found from
the integral over the \textbf{k}--space:%
\begin{equation}
\rho (\mathbf{r})=T\sum_{i\omega }\sum_{\mathbf{k}j}\sum_{\sigma }g_{\mathbf{%
k}j\omega }\psi _{\mathbf{k}j\omega }^{L(\sigma )}(\mathbf{r})\psi _{\mathbf{%
k}j\omega }^{R(\sigma )}(\mathbf{r})e^{i\omega 0^{+}}  \label{IMPeksRHO}
\end{equation}

\begin{equation}
\mathbf{m}(\mathbf{r})=g\mu _{B}T\sum_{i\omega }\sum_{\mathbf{k}j}g_{\mathbf{%
k}j\omega }\sum_{\sigma \sigma ^{\prime }}\mathbf{s}_{\sigma \sigma ^{\prime
}}\psi _{\mathbf{k}j\omega }^{L(\sigma )}(\mathbf{r})\psi _{\mathbf{k}%
j\omega }^{R(\sigma ^{\prime })}(\mathbf{r})e^{i\omega 0^{+}}
\label{IMPeksMAG}
\end{equation}

The central procedure which is yet to be determined is the construction of
the exchange--correlation part of the self--energy. It is formally a
variational derivative of the exchange correlation part of the free energy
functional%
\begin{equation}
\mathcal{M}_{xc}^{\sigma \sigma ^{\prime }}(\mathbf{r},\mathbf{r}^{\prime
},i\omega )=\frac{\delta \Phi _{SDF}^{xc}[G_{loc}]}{\delta \mathcal{G}%
^{\sigma ^{\prime }\sigma }(\mathbf{r}^{\prime },\mathbf{r},i\omega )}=\frac{%
\delta \Phi _{SDF}^{xc}[G_{loc}]}{\delta G_{loc}^{\sigma ^{\prime }\sigma }(%
\mathbf{r}^{\prime },\mathbf{r},i\omega )}\theta _{loc}(\mathbf{r},\mathbf{r}%
^{\prime })  \label{IMPeksMXC}
\end{equation}%
The spectral density functional theory, where an exact functional of \
certain local quantities is constructed in the spirit of Ref. %
\onlinecite{Chitra1} uses effective self--energies which are local by
construction. This property can be exploited to find good approximations to
the interaction energy functional. For example, if it is \textit{a priori}
known that the real electronic self--energy is local in a certain portion of
the Hilbert space, a good approximation is the corresponding local dynamical
mean field theory obtained for example by a restriction or truncation of the
full Baym--Kadanoff functional to local quantities in the spirit of Ref. %
\onlinecite{Chitra2}.The local DMFT approximates the functional $\Phi
_{SDF}[G_{loc}]$ by the sum of all two--particle diagrams evaluated with $%
G_{loc}$ and the bare Coulomb interaction $v_{C}$ $.$ In other words, the
functional dependence of the interaction part $\Phi _{BK}[G]$ in the
Baym--Kadanoff functional for which the diagrammatic rules exist is now
restricted by $G_{loc}$ and is used as an approximation to $\Phi
_{SDF}[G_{loc}]$, i.e. $\Phi _{SDF}[G_{loc}]=\Phi _{BK}[G_{loc}].$ Obviously
that the variational derivative of such restricted functional will generate
the self--energy confined in the same area as the local Green function
itself.

Remarkably the summation over all local diagrams can be performed exactly
via introduction of an auxiliary quantum impurity model subjected to a
self--consistency condition \cite{DMFTReview}. If this impurity is
considered as a cluster $C$, the cellular DMFT (C--DMFT) can be used which
breaks the translational invariance of the lattice to obtain accurate
estimates of the self energies.

Various methods such as LDA+U \cite{LDA+U}, LDA+DMFT\cite{ReviewLDA+DMFT}
and local GW\cite{Zein,Tsvelik} which appeared recently for realistic
calculations of properties of strongly correlated materials can be naturally
understood within spectral density functional theory. Let us, for example,
explore the idea of expressing the energy as the density functional. Local
density approximation prompts us that a large portion of the
exchange--correlation part $\Phi _{xc}[\rho ]$ can be found easily. Indeed,
the charge density is known to be accurately obtained by the LDA. Why not
think of LDA as the most primitive impurity solver, which generates
manifestly local self--energy with localization radius collapsed to a single 
$\mathbf{r}$ point? It is tempting to represent $\Phi
_{SDF}[G_{loc}]=E_{H}[\rho ]+E_{xc}^{LDA}[\rho ]+\tilde{\Phi}[G_{loc}]-\Phi
_{DC}[G_{loc}],$ where the new functional $\tilde{\Phi}_{SDF}[G_{loc}]$
needs in fact to take care of those electrons which are strongly correlated
and heavy, thus badly described by LDA. Conceptually, that means that the
solution of the cluster impurity model for the light electrons is
approximated by LDA and does not need a frequency resolution for their
self--energies.

Such introduced LDA+DMFT\ approximation considers both the density and the
local Green function as the parameters of the spectral density functional .
A further approximation is made to accelerate the solution of a single--site
impurity model: the functional dependence comes from the subblock of the
correlated electrons only. If localized orbital representation $\{\chi
_{\alpha }\}$ is utilized, a subspace of the heavy electrons $\{\chi _{a}\}$
can be identified. Thus, the approximation can be written as $\tilde{\Phi}%
_{SDF}[G_{loc,ab}(i\omega )]$ , where $G_{loc,ab}(i\omega )$ is the heavy
block of the local Green function$.$

Unfortunately, the LDA has no diagrammatic representation, and it is
difficult to separate the contributions from the light and heavy electrons.
The $E_{xc}^{LDA}[\rho ]$ is a non--linear functional and it already
includes the contribution to the energy from all orbitals in some average
form. Therefore we need to take care of a non--trivial double counting,
encoded in the functional $\Phi _{DC}[G_{loc}]$. The precise form of the
double counting is related to the approximation imposed for $\tilde{\Phi}%
[G_{loc}]$ and can be borrowed from the LDA+U method \cite{LDA+U}.  Note
that both LDA+U and LDA+DMFT methods require separations of the electrons
onto light and heavy which makes them\ basis--set dependent.

Iron and nickel were studied in Refs. %
\onlinecite{Lichtenstein:2000,Lichtenstein:2001}. The values $U=2.3$ $(3.0)$%
~eV for Fe (Ni) and the same value of the interatomic exchange, $J=0.9$~eV
for both Fe and Ni were used, as came out from the constrained LDA
calculations \cite{LDA++,Bandyopadhyay:1989,Lichtenstein:1997}. These
parameters are consistent with those of many earlier studies and resulted in
a good description of the physical properties of Fe and Ni. In Ref. %
\onlinecite{Lichtenstein:2001} the general form of the double counting
correction $V_{\sigma }^{DC}=\frac{1}{2}Tr_{\sigma }\mathcal{M}_{\sigma }(0)$
was taken. Notice that because of the different self energies in the $e_{g}$
and $t_{2g}$ blocks the DMFT Fermi surface does not coincide with the LDA
Fermi surface.

The impurity model was solved by QMC in Ref.~ \onlinecite{Lichtenstein:2001}
and by the FLEX scheme in Ref. \onlinecite{Katsnelson:2002}. It is clear
that nickel is more itinerant than iron (the spin--spin autocorrelation
decays faster), which has longer lived spin fluctuations. On the other hand,
the one--particle density of states of iron resembles very much the LSDA
density of states while the DOS of nickel below $T_{c}$, has additional
features which are not present in the LSDA spectra \cite{Iwan:1979,
Eberhardt:1980, Altmann:2000}: the presence of a famous 6~eV satellite, the
30\% narrowing of the occupied part of $d$-band and the 50\% decrease of
exchange spittoons compared to the LDA results. Note that the satellite in
Ni has substantially more spin-up contributions in agreement with
photoemission spectra \cite{Altmann:2000}. The exchange splitting of the
d--band depends very weakly on temperature from $T=0.6T_{C}$ to $T=0.9T_{C}$%
. Correlation effects in Fe are less pronounced than in Ni, due to its large
spin splitting and the characteristic bcc structural dip in the density of
states for the spin--down states near the Fermi level, which reduces the
density of states for particle hole excitations.

\begin{figure}[tbp]
\epsfig{file=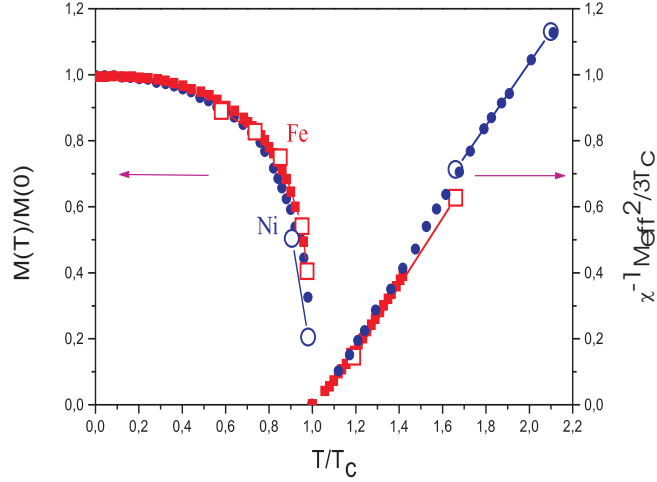, width=9cm,height=7cm}
\caption{Temperature dependence of ordered moment and the inverse
ferromagnetic susceptibility for Fe (open square) and Ni (open circle)
compared with experimental results for Fe (square) and Ni (circle) (from
Ref.~\onlinecite{Stocks:1998}). The calculated moments were normalized to
the LDA ground state magnetization (2.2 $\protect\mu _{B}$ for Fe and 0.6 $%
\protect\mu _{B}$ for Ni). }
\label{figMOMFe}
\end{figure}

The uniform spin susceptibility in the paramagnetic state, ${\chi }%
_{q=0}=dM/dH$, was extracted from the QMC simulations by measuring the
induced magnetic moment in a small external magnetic field. It includes the
polarization of the impurity Weiss field by the external field \cite%
{DMFTReview}. The dynamical mean field results account for the Curie--Weiss
law which is observed experimentally in Fe and Ni. As the temperature
increases above $T_{c}$, the atomic character of the system is partially
restored resulting in an atomic like susceptibility with an effective
moment: 
\begin{equation}
\chi _{q=0}=\frac{\mu _{eff}^{2}}{3(T-T_{C})}.  \label{eq:effective}
\end{equation}%
The temperature dependence of the ordered magnetic moment below the Curie
temperature and the inverse of the uniform susceptibility above the Curie
point are plotted in Fig.~\ref{figMOMFe} together with the corresponding
experimental data for iron and nickel \cite{Wolfarth:1986}. The LDA+DMFT
calculation describes the magnetization curve and the slope of the
high--temperature Curie--Weiss susceptibility remarkably well. The
calculated values of high--temperature magnetic moments extracted from the
uniform spin susceptibility are $\mu _{eff}=3.09$ $(1.50)\mu _{B}$ for Fe
(Ni), in good agreement with the experimental data $\mu _{eff}=3.13$ $%
(1.62)\mu _{B}$ for Fe (Ni)~\cite{Wolfarth:1986}.

The Curie temperatures of Fe and Ni were estimated from the disappearance of
spin polarization in the self--consistent solution of the DMFT problem and
from the Curie--Weiss law in Eq.(\ref{eq:effective}). The estimates for $%
T_{C}=1900$ $(700)$~K are in reasonable agreement with experimental values
of $1043$ $(631)$~K for Fe (Ni) respectively \cite{Wolfarth:1986},
considering the single--site nature of the DMFT approach, which is not able
to capture the reduction of $T_{C}$ due to long--wavelength spin waves.
These effects are governed by the spin--wave stiffness. Since the ratio of
the spin--wave stiffness $D$ to $T_{C}$, $T_{C}$/$a^{2}D,$ is nearly a
factor of 3 larger for Fe than for Ni \cite{Wolfarth:1986} ($a$ is the
lattice constant), we expect the $T_{C}$ in DMFT to be much higher than the
observed Curie temperature in Fe than in Ni. Quantitative calculations
demonstrating the sizeable reduction of $T_{C}$ due to spin waves in Fe in
the framework of a Heisenberg model were performed in Ref. \onlinecite{Pajda}%
. This physics whereby the long wavelength fluctuations renormalize the
critical temperature would be reintroduced in the DMFT by the use E--DMFT.
Alternatively, the reduction of the critical temperature due to spatial
fluctuations can be investigated with cluster DMFT methods.

Within the dynamical mean field theory one can also compute the local spin
susceptibility defined by 
\begin{equation}
\chi _{loc}=\frac{g^{2}}{3}\int\limits_{0}^{\beta }d\tau \left\langle 
\mathbf{S}\left( \tau \right) \mathbf{S}(0)\right\rangle
\label{eq:chi_local}
\end{equation}%
where $\mathbf{S}=\frac{1}{2}\sum_{m,\sigma ,\sigma ^{\prime }}c_{m\sigma
}^{\dagger }\mathbf{\sigma }_{\sigma \sigma ^{\prime }}c_{m\sigma ^{\prime
}} $ is single-site spin operator. It differs from the $q=0$ susceptibility
by the absence of spin polarization in the Weiss field of the impurity
model. Eq.(\ref{eq:chi_local}) cannot be probed directly in experiments but
it is easily computed within the DMFT--QMC. Its behavior as a function of
temperature gives a very intuitive picture of the degree of correlations in
the system. In a weakly correlated regime we expect Eq.(\ref{eq:chi_local})
to be nearly temperature independent, while in a strongly correlated regime
we expect a leading Curie--Weiss behavior at high temperatures $\chi
_{local}=\mu _{loc}^{2}/({3}T+const.)$ where $\mu _{loc}$ is an effective
local magnetic moment. In the Heisenberg model with spin $S$, $\mu
_{loc}^{2}=S(S+1)g^{2}$ and for the well--defined local magnetic moments
(e.g., for rare--earth magnets) this quantity should be temperature
independent. For the itinerant electron magnets, $\mu _{loc}$ is temperature
dependent due to a variety of competing many--body effects such as Kondo
screening, the induction of local magnetic moment by temperature \cite{Moria}
and thermal fluctuations which disorder the moments \cite{Irkhin:1994}. All
these effects are included in the DMFT calculations.

The comparison of the values of the local and the $q=0$ susceptibility gives
a crude measure of the degree of short--range order which is present above $%
T_{C}$. As expected, the moments extracted from the local susceptibility Eq.(%
\ref{eq:chi_local}) are a bit smaller (2.8 \ $\mu _{B}$ for iron and 1.3 \ $%
\mu _{B}$ for nickel) than those extracted from the uniform magnetic
susceptibility. This reflects the small degree of the short--range
correlations which remain well above $T_{C}$ \cite{Mook:1985}. The
high--temperature LDA+DMFT clearly shows the presence of a local moment
above $T_{C}$. This moment is correlated with the presence of high--energy
features (of the order of the Coulomb energies) in the photoemission. This
is also true below $T_{C}$, where the spin dependence of the spectra is more
pronounced for the satellite region in Nickel than for that of the
quasiparticle bands near the Fermi level. This can explain the apparent
discrepancies between different experimental determinations of the
high--temperature magnetic splittings \cite{Kisker:1984, Kakizaki:1994,
Sinkovic:1997, Kreutz:1998} as being the results of probing different energy
regions. The resonant photoemission experiments \cite{Sinkovic:1997} reflect
the presence of local--moment polarization in the high--energy spectrum
above the Curie temperature in Nickel, while the low--energy angle resolved
photoemission investigations \cite{Kreutz:1998} results in non--magnetic
bands near the Fermi level. This is exactly the DMFT view on the electronic
structure of transition metals above $T_{C}$. Fluctuating moments and
atomic--like configurations are large at short times, which results in
correlation effects in the high--energy spectra such as spin--multiplet
splittings. The moment is reduced at longer time scales, corresponding to a
more band--like, less correlated electronic structure near the Fermi level.

NiO and MnO represent two classical Mott--Hubbard systems. Both materials
are insulators with the energy gap of a few eV regardless whether they are
antiferro-- or paramagnetic. The spin dependent LSDA theory strongly
underestimates the energy gap in the ordered phase. This can be corrected by
the use of the LDA+U\ method. Both theories however fail completely to
describe the local moment regime reflecting a general drawback of band
theory to reproduce atomic limit. Therefore the real challenge is to
describe the paramagnetic insulating state where the self--energy effects
are crucial both for the electronic structure and for recovering the correct
phonon dispersions in these materials. The DMFT calculations have been
performed \cite{Savrasov:2003} by taking into account correlations among $d$%
--electrons. In the regime of large U adequate for both for NiO and MnO in
the paramagnetic phase the correlations were treated within the well--known
Hubbard~I approximation.

The calculated densities of states using the LDA+DMFT method for the
paramagnetic state of NiO and MnO \cite{Savrasov:2003} have revealed the
presence of both lower and upper Hubbard subbands. These were found in
agreement with the LDA+U calculations of Anisimov \cite{LDA+U} which have
been performed for the ordered states of these oxides. Clearly, spin
integrated spectral functions do not show an appreciable dependence with
temperature and look similar below and above phase transition point.

The progress in the electronic structure calculation with DMFT allowed to
apply the method to more complicated systems such as plutonium \cite%
{Nature,Science}. The temperature dependence of atomic volume in Pu is
anomalous \cite{PuBook,PuVol}. It shows an enormous volume expansion between 
$\alpha $ and $\delta $ phases which is about 25\%. Within the $\delta $
phase, the metal shows negative thermal expansion. Transition between $%
\delta $ and higher--temperature $\varepsilon $ phase occurs with a 5\%
volume collapse. Also, Pu shows anomalous resistivity behavior \cite{PuRHO}
characteristic for the heavy fermion systems, but neither of its phases is
magnetic. The susceptibility is small and relatively temperature
independent. The photoemission \cite{PuExp1} shows a strong narrow
Kondo--like peak at the Fermi level consistent with large values of the
linear specific heat coefficient.

Density functional based LDA\ and GGA\ calculations describe the properties
of Pu incorrectly. They predict magnetic ordering \cite{PuMag}. They
underestimate \cite{PuLDA} equilibrium volume of the $\delta $ and $%
\varepsilon $ phase by as much as 30\%, the largest discrepancy known in
LDA, which usually predicts the volume of solids within a few \% accuracy
even for such correlated systems as high temperature superconductors.

To address these questions several approaches have been developed. The LDA+U
method was applied to $\delta $ Pu \cite{PuPRL}. It is able to produce the
correct volume of the $\delta $ phase for values of the parameter $U\symbol{%
126}$4 eV consistent with atomic spectral data and constrained density
functional calculations. Similar calculation has been performed by a
so--called orbitally ordered density functional method \cite{PuLDA+J}.
However, both methods predict Pu to be magnetic, which is not observed
experimentally. The LDA+U method is unable to predict the correct excitation
spectrum. Also, to recover the $\alpha $ phase within LDA+U$\ $the parameter
U has to be set to zero which is inconsistent with its transport properties
and with microscopic calculations of this parameter. Another approach
proposed \cite{PuCore} in the past is the constrained LDA approach in which
some of the 5f electrons, are treated as core, while the remaining are
allowed to participate in band formation. Results of the
self--interaction--corrected LDA calculations have been reported \cite{PuSIC}%
, as well as qualitative discussion of the bonding nature across the
actinides series has been given \cite{Harrison}.

Dynamical mean--field calculations have been recently applied with success
to this material. They were performed in paramagnetic state and do not
impose magnetic ordering. DMFT was able to predict the appearance of a
strong quasiparticle peak near the Fermi level which exists in the both
phases. Also, the lower and upper Hubbard bands were clearly distinguished.
The width of the quasiparticle peak in the $\alpha $ phase is found to be
larger by 30 per cent compared to the width in the $\delta $ phase. This
indicates that the low--temperature phase is more metallic, i.e. it has
larger spectral weight in the quasiparticle peak and smaller weight in the
Hubbard bands. Recent advances have allowed the experimental determination
of these spectra, and these calculations were consistent with these
measurements \cite{PuExp1}.

\section{Conclusion}

In conclusion, electronic structure methods have been applied with success
to study magnetic properties of solids. Many good results have been obtained
for weakly correlated itinerant systems. New algorithms for inclusion of
many--body effects are beginning to attack magnetic phenomena in strongly
correlated materials. These successes have generated a new synergy between
the electronic structure and many--body communities, which are jointly
developing new and powerful electronic structure tools for virtual material
exploration.

The work on magnetic anisotropy was supported by the Office of Naval
Research grant No. 4-2650.

\end{document}